\input harvmac
\noblackbox

\input epsf

\newcount\figno
\figno=0
\def\fig#1#2#3{
\par\begingroup\parindent=0pt\leftskip=1cm\rightskip=1cm\parindent=0pt
\baselineskip=11pt
\global\advance\figno by 1
\midinsert
\epsfxsize=#3
\centerline{\epsfbox{#2}}
\vskip 12pt
{\bf Fig.\ \the\figno: } #1\par
\endinsert\endgroup\par
}
\def\figlabel#1{\xdef#1{\the\figno}}
\def\encadremath#1{\vbox{\hrule\hbox{\vrule\kern8pt\vbox{\kern8pt
\hbox{$\displaystyle #1$}\kern8pt}
\kern8pt\vrule}\hrule}}

\def\apm{{\alpha^{\prime}}}

\font\cmss=cmss10
\font\cmsss=cmss10 at 7pt
\def\rlx{\relax\leavevmode}
\def\inbar{\vrule height1.5ex width.4pt depth0pt}
\def\IN{\relax{\rm I\kern-.18em N}}
\def\IP{\relax{\rm I\kern-.18em P}}
\def\ZZ{\rlx\leavevmode\ifmmode\mathchoice{\hbox{\cmss Z\kern-.4em Z}}
 {\hbox{\cmss Z\kern-.4em Z}}{\lower.9pt\hbox{\cmsss Z\kern-.36em Z}}
 {\lower1.2pt\hbox{\cmsss Z\kern-.36em Z}}\else{\cmss Z\kern-.4em
 Z}\fi}
\def\IZ{\relax\ifmmode\mathchoice
{\hbox{\cmss Z\kern-.4em Z}}{\hbox{\cmss Z\kern-.4em Z}}
{\lower.9pt\hbox{\cmsss Z\kern-.4em Z}}
{\lower1.2pt\hbox{\cmsss Z\kern-.4em Z}}\else{\cmss Z\kern-.4em
Z}\fi}
\def\IZ{\relax\ifmmode\mathchoice
{\hbox{\cmss Z\kern-.4em Z}}{\hbox{\cmss Z\kern-.4em Z}}
{\lower.9pt\hbox{\cmsss Z\kern-.4em Z}}
{\lower1.2pt\hbox{\cmsss Z\kern-.4em Z}}\else{\cmss Z\kern-.4em
Z}\fi}

\def\narrowplus{\kern -.04truein + \kern -.03truein}
\def\narrowminus{- \kern -.04truein}
\def\narrowminussub{\kern -.02truein - \kern -.01truein}

\def\t{{\theta}}
\def\l{{\lambda}}

\def\frac#1#2{{#1\over #2}}

\def\CM{{\cal M}}
\def\IZ{\relax\ifmmode\mathchoice
{\hbox{\cmss Z\kern-.4em Z}}{\hbox{\cmss Z\kern-.4em Z}}
{\lower.9pt\hbox{\cmsss Z\kern-.4em Z}}
{\lower1.2pt\hbox{\cmsss Z\kern-.4em Z}}\else{\cmss Z\kern-.4em
Z}\fi}
\def\IC{{\relax\,\hbox{$\inbar\kern-.3em{\rm C}$}}}
\def\p{\partial}
\font\cmss=cmss10 \font\cmsss=cmss10 at 7pt
\def\IR{\relax{\rm I\kern-.18em R}}
\def\ra{\rangle}
\def\la{\langle}

\def\half{{1\over 2}}

\def\IZ{\relax\ifmmode\mathchoice
{\hbox{\cmss Z\kern-.4em Z}}{\hbox{\cmss Z\kern-.4em Z}}
{\lower.9pt\hbox{\cmsss Z\kern-.4em Z}}
{\lower1.2pt\hbox{\cmsss Z\kern-.4em Z}}\else{\cmss Z\kern-.4em
Z}\fi}
\def\IC{{\relax\,\hbox{$\inbar\kern-.3em{\rm C}$}}}

\def\b{{\beta}}

\def\a{{\alpha}}

\def\D{{\Delta}}
\def\m{{\mu}}
\def\n{{\nu}}
\def\ep{{\epsilon}}
\def\epp{{\epsilon^{\prime}}}
\def\d{{\delta}}

\def\G{{\Gamma}}

\def\t{{\theta}}
\def\l{{\lambda}}

\def\P{{\Phi}}

\def\t{{\tau}}
\def\r{{\rho}}
\def\th{{\theta}}

\def\R{{\rightarrow}}

\def\frac#1#2{{#1\over #2}}

\def\CM{{\cal M}}
\def\CN{{\cal N}}

\def\p{\partial}

\def\apm{\alpha^{\prime}}

\lref\th{G.~'tHooft, ``A Planar Diagram Theory for Strong Interactions,'' Nucl.\ Phys.\ 
{\bf 72}, 461, (1974).}
\lref\gv{
R.~Gopakumar and C.~Vafa,
``On the gauge theory/geometry correspondence,''
Adv.\ Theor.\ Math.\ Phys.\  {\bf 3}, 1415 (1999)
[arXiv:hep-th/9811131].}
\lref\ov{
H.~Ooguri and C.~Vafa,
``Worldsheet derivation of a large N duality,''
Nucl.\ Phys.\ B {\bf 641}, 3 (2002)
[arXiv:hep-th/0205297].}
\lref\malda{
J.~M.~Maldacena,
``The large $N$ limit of superconformal field theories and supergravity,''
Adv.\ Theor.\ Math.\ Phys.\  {\bf 2}, 231 (1998)
[Int.\ J.\ Theor.\ Phys.\  {\bf 38}, 1113 (1999)]
[arXiv:hep-th/9711200].}
\lref\divec{P.~Di Vecchia, L.~Magnea, A.~Lerda, R.~Russo and R.~Marotta,
``String techniques for the calculation of renormalization constants in field theory,''
Nucl.\ Phys.\ B {\bf 469}, 235 (1996)
[arXiv:hep-th/9601143].}
\lref\fmr{A.~Frizzo, L.~Magnea and R.~Russo,
``Systematics of one-loop Yang-Mills diagrams from bosonic string  amplitudes,''
Nucl.\ Phys.\ B {\bf 604}, 92 (2001)
[arXiv:hep-ph/0012129].}
\lref\berkos{
Z.~Bern and D.~A.~Kosower,
``Efficient Calculation Of One Loop QCD Amplitudes,''
Phys.\ Rev.\ Lett.\  {\bf 66}, 1669 (1991).}
\lref\bdh{
L.~Brink, P.~Di Vecchia and P.~S.~Howe,
``A Lagrangian Formulation Of The Classical And Quantum Dynamics Of Spinning Particles,''
Nucl.\ Phys.\ B {\bf 118}, 76 (1977).}
\lref\ssen{M.~Tuite and S.~Sen,
``A String Motivated Approach to the Relativistic Point Particle,''
arXiv:hep-th/0308099.}
\lref\strass{M.~J.~Strassler,
``Field theory without Feynman diagrams: One loop effective actions,''
Nucl.\ Phys.\ B {\bf 385}, 145 (1992)
[arXiv:hep-ph/9205205].}
\lref\schb{C.~Schubert,
``Perturbative quantum field theory in the string-inspired formalism,''
Phys.\ Rept.\  {\bf 355}, 73 (2001)
[arXiv:hep-th/0101036].}
\lref\mikh{A.~Mikhailov,
``Notes on higher spin symmetries,''
arXiv:hep-th/0201019.}
\lref\cnss{G.~Chalmers, H.~Nastase, K.~Schalm and R.~Siebelink,
``R-current correlators in N = 4 super Yang-Mills theory from anti-de  Sitter supergravity,''
Nucl.\ Phys.\ B {\bf 540}, 247 (1999)
[arXiv:hep-th/9805105].}
\lref\guil{E. A. Guillemin, ``Introductory Circuit Theory'', (John Wiley and Sons, 1953).}
\lref\BD{J. D. Bjorken and S. D. Drell, ``Relativistic Quantum Fields'', (McGraw Hill, 1965).}
\lref\lam{C.~S.~Lam,
``Multiloop string - like formulas for QED,''
Phys.\ Rev.\ D {\bf 48}, 873 (1993)
[arXiv:hep-ph/9212296].}
\lref\petkou{A.~Petkou,
``Conserved currents, consistency relations, and operator product  expansions 
in the conformally invariant O(N) vector model,''
Annals Phys.\  {\bf 249}, 180 (1996)
[arXiv:hep-th/9410093].}
\lref\liu{H.~Liu,
``Scattering in anti-de Sitter space and operator product expansion,''
Phys.\ Rev.\ D {\bf 60}, 106005 (1999)
[arXiv:hep-th/9811152].}
\lref\dmmr{E.~D'Hoker, S.~D.~Mathur, A.~Matusis and L.~Rastelli,
``The operator product expansion of N = 4 SYM and the 4-point functions  of supergravity,''
Nucl.\ Phys.\ B {\bf 589}, 38 (2000)
[arXiv:hep-th/9911222].}
\lref\dfmmr{E.~D'Hoker, D.~Z.~Freedman, S.~D.~Mathur, A.~Matusis and L.~Rastelli,
``Graviton exchange and complete 4-point functions in the AdS/CFT  correspondence,''
Nucl.\ Phys.\ B {\bf 562}, 353 (1999)
[arXiv:hep-th/9903196].}
\lref\fmmr{D.~Z.~Freedman, S.~D.~Mathur, A.~Matusis and L.~Rastelli,
``Correlation functions in the CFT($d$)/AdS($d+1$) correspondence,''
Nucl.\ Phys.\ B {\bf 546}, 96 (1999)
[arXiv:hep-th/9804058].}
\lref\sund{P.~Haggi-Mani and B.~Sundborg,
``Free large N supersymmetric Yang-Mills theory as a string theory,''
JHEP {\bf 0004}, 031 (2000)
[arXiv:hep-th/0002189];}
\lref\sundb{B.~Sundborg,
``Stringy gravity, interacting tensionless strings and massless higher  spins,''
Nucl.\ Phys.\ Proc.\ Suppl.\  {\bf 102}, 113 (2001)
[arXiv:hep-th/0103247].}
\lref\witt{E. Witten, Talk at the John Schwarz 60th Birthday Symposium,
http://theory.caltech.edu/jhs60/witten.}
\lref\dmw{A.~Dhar, G.~Mandal and S.~R.~Wadia,
``String bits in small radius AdS and weakly coupled N = 4 super  Yang-Mills theory. I,''
arXiv:hep-th/0304062.}
\lref\dnw{L.~Dolan, C.~R.~Nappi and E.~Witten,
``A relation between approaches to integrability in superconformal Yang-Mills theory,''
arXiv:hep-th/0308089.}
\lref\polch{J.~Polchinski, Seminar at SW workshop on String Theory (Feb. 2003), unpublished.}
\lref\msw{G.~Mandal, N.~V.~Suryanarayana and S.~R.~Wadia,
``Aspects of semiclassical strings in AdS(5),''
Phys.\ Lett.\ B {\bf 543}, 81 (2002)
[arXiv:hep-th/0206103].}
\lref\bnp{I.~Bena, J.~Polchinski and R.~Roiban,
``Hidden symmetries of the AdS(5) x S**5 superstring,''
arXiv:hep-th/0305116.}
\lref\vall{B.~C.~Vallilo,
``Flat currents in the classical AdS(5) x S**5 pure spinor superstring,''
arXiv:hep-th/0307018.}
\lref\klepol{I.~R.~Klebanov and A.~M.~Polyakov,
``AdS dual of the critical O(N) vector model,''
Phys.\ Lett.\ B {\bf 550}, 213 (2002)
[arXiv:hep-th/0210114].}
\lref\peta{A.~C.~Petkou,
``Evaluating the AdS dual of the critical O(N) vector model,''
JHEP {\bf 0303}, 049 (2003)
[arXiv:hep-th/0302063].}
\lref\leigh{R.~G.~Leigh and A.~C.~Petkou,
``Holography of the N = 1 higher-spin theory on AdS(4),''
JHEP {\bf 0306}, 011 (2003)
[arXiv:hep-th/0304217].}
\lref\sezsuna{E.~Sezgin and P.~Sundell,
``Holography in 4D (super) higher spin theories and a test via cubic  scalar couplings,''
arXiv:hep-th/0305040.}
\lref\por{L.~Girardello, M.~Porrati and A.~Zaffaroni,
``3-D interacting CFTs and generalized Higgs phenomenon in higher spin  theories on AdS,''
Phys.\ Lett.\ B {\bf 561}, 289 (2003)
[arXiv:hep-th/0212181].}
\lref\ruhl{T.~Leonhardt, A.~Meziane and W.~Ruhl,
``On the proposed AdS dual of the critical O(N) sigma model for any  dimension $2 < d < 4$,''
Phys.\ Lett.\ B {\bf 555}, 271 (2003)
[arXiv:hep-th/0211092].}
\lref\rajan{F.~Kristiansson and P.~Rajan,
``Scalar field corrections to AdS(4) gravity from higher spin gauge  theory,''
JHEP {\bf 0304}, 009 (2003)
[arXiv:hep-th/0303202].}
\lref\sumit{S.~R.~Das and A.~Jevicki,
``Large-N collective fields and holography,''
Phys.\ Rev.\ D {\bf 68}, 044011 (2003)
[arXiv:hep-th/0304093].}
\lref\nemani{N.~V.~Suryanarayana,
JHEP {\bf 0306}, 036 (2003)
[arXiv:hep-th/0304208].}
\lref\pol{A.~M.~Polyakov,
``Gauge fields and space-time,''
Int.\ J.\ Mod.\ Phys.\ A {\bf 17S1}, 119 (2002)
[arXiv:hep-th/0110196].}
\lref\polb{A.~M.~Polyakov,
``String theory and quark confinement,''
Nucl.\ Phys.\ Proc.\ Suppl.\  {\bf 68}, 1 (1998)
[arXiv:hep-th/9711002].}
\lref\polc{A.~M.~Polyakov,
``The wall of the cave,''
Int.\ J.\ Mod.\ Phys.\ A {\bf 14}, 645 (1999)
[arXiv:hep-th/9809057].}
\lref\polbk{A.~M.~Polyakov,
``Gauge fields and Strings,'' (Harwood Academic Publishers, 1987).} 
\lref\bmn{D.~Berenstein, J.~M.~Maldacena and H.~Nastase,
``Strings in flat space and pp waves from N = 4 super Yang Mills,''
JHEP {\bf 0204}, 013 (2002)
[arXiv:hep-th/0202021].}
\lref\gkp{S.~S.~Gubser, I.~R.~Klebanov and A.~M.~Polyakov,
``Gauge theory correlators from non-critical string theory,''
Phys.\ Lett.\ B {\bf 428}, 105 (1998)
[arXiv:hep-th/9802109].}
\lref\witads{E.~Witten,
``Anti-de Sitter space and holography,''
Adv.\ Theor.\ Math.\ Phys.\  {\bf 2}, 253 (1998)
[arXiv:hep-th/9802150].}
\lref\sezsun{E.~Sezgin and P.~Sundell,
``Doubletons and 5D higher spin gauge theory,''
JHEP {\bf 0109}, 036 (2001)
[arXiv:hep-th/0105001]; ``Massless higher spins and holography,''
Nucl.\ Phys.\ B {\bf 644}, 303 (2002)
[Erratum-ibid.\ B {\bf 660}, 403 (2003)]
[arXiv:hep-th/0205131].}
\lref\vas{
M.~A.~Vasiliev,
``Conformal higher spin symmetries of 4D massless supermultiplets and  osp(L,2M) 
invariant equations in generalized (super)space,''
Phys.\ Rev.\ D {\bf 66}, 066006 (2002)
[arXiv:hep-th/0106149].}
\lref\vasrev{M.~A.~Vasiliev,
``Higher spin gauge theories: Star-product and AdS space,''
arXiv:hep-th/9910096.}
\lref\bv{A.~O.~Barvinsky and G.~A.~Vilkovisky,
``Beyond The Schwinger-Dewitt Technique: Converting Loops Into Trees And In-In Currents,''
Nucl.\ Phys.\ B {\bf 282}, 163 (1987).}
\lref\birdav{N.~D.~Birrell and P.~C.~W.~Davies, ``Quantum fields in Curved Space,'' Cambridge University
Press (1984).}
\lref\hsken{M.~Henningson and K.~Skenderis,
``The holographic Weyl anomaly,''
JHEP {\bf 9807}, 023 (1998)
[arXiv:hep-th/9806087].}
\lref\nojod{S.~Nojiri and S.~D.~Odintsov,
``Conformal anomaly for dilaton coupled theories from AdS/CFT  correspondence,''
Phys.\ Lett.\ B {\bf 444}, 92 (1998)
[arXiv:hep-th/9810008].}
\lref\odnoj{S.~Nojiri, S.~D.~Odintsov and S.~Ogushi,
``Finite action in d5 gauged supergravity and dilatonic conformal anomaly  for dual quantum field theory,''
Phys.\ Rev.\ D {\bf 62}, 124002 (2000)
[arXiv:hep-th/0001122].}
\lref\dss{S.~de Haro, S.~N.~Solodukhin and K.~Skenderis,
``Holographic reconstruction of spacetime and renormalization in the  AdS/CFT correspondence,''
Commun.\ Math.\ Phys.\  {\bf 217}, 595 (2001)
[arXiv:hep-th/0002230].}
\lref\feffgr{C.~Fefferman, C.~R.~Graham, ``Conformal Invariants'' in ``{\it Elie Cartan et les 
Mathematiques d'Aujourd'hui},'' (Asterisque 1985), 95.}
\lref\glebfrpet{G.~Arutyunov, S.~Frolov and A.~C.~Petkou,
``Operator product expansion of the lowest weight CPOs in N = 4  SYM(4) at strong coupling,''
Nucl.\ Phys.\ B {\bf 586}, 547 (2000)
[Erratum-ibid.\ B {\bf 609}, 539 (2001)]
[arXiv:hep-th/0005182]; 
``Perturbative and instanton corrections to the OPE of CPOs in N = 4  SYM(4),''
Nucl.\ Phys.\ B {\bf 602}, 238 (2001)
[Erratum-ibid.\ B {\bf 609}, 540 (2001)]
[arXiv:hep-th/0010137].}
\lref\arutfrol{G.~Arutyunov and S.~Frolov,
``Three-point Green function of the stress-energy tensor in the AdS/CFT  correspondence,''
Phys.\ Rev.\ D {\bf 60}, 026004 (1999)
[arXiv:hep-th/9901121].}
\lref\east{M.~G.~Eastwood,
``Higher symmetries of the Laplacian,''
arXiv:hep-th/0206233.}
\lref\gm{D.~J.~Gross and P.~F.~Mende,
``String Theory Beyond The Planck Scale,''
Nucl.\ Phys.\ B {\bf 303}, 407 (1988).}
\lref\gross{D.~J.~Gross,
``High-Energy Symmetries Of String Theory,''
Phys.\ Rev.\ Lett.\  {\bf 60}, 1229 (1988).}
\lref\bianchi{M.~Bianchi, J.~F.~Morales and H.~Samtleben,
``On stringy AdS(5) x S**5 and higher spin holography,''
JHEP {\bf 0307}, 062 (2003).}
\lref\anselmi{D.~Anselmi,
``The N = 4 quantum conformal algebra,''
Nucl.\ Phys.\ B {\bf 541}, 369 (1999)
[arXiv:hep-th/9809192].}
\lref\lind{U.~Lindstrom and M.~Zabzine,
``Tensionless strings, WZW models at critical level and massless higher  spin fields,''
arXiv:hep-th/0305098.}
\lref\ashoke{A.~Sen,
``Open-closed duality at tree level,''
arXiv:hep-th/0306137.}
\lref\gir{D.~Gaiotto, N.~Itzhaki and L.~Rastelli,
``Closed strings as imaginary D-branes,''
arXiv:hep-th/0304192.}
\lref\satchi{S.~R.~Das, S.~Naik and S.~R.~Wadia,
``Quantization Of The Liouville Mode And String Theory,''
Mod.\ Phys.\ Lett.\ A {\bf 4}, 1033 (1989).}
\lref\spenta{A.~Dhar, T.~Jayaraman, K.~S.~Narain and S.~R.~Wadia,
``The Role Of Quantized Two-Dimensional Gravity In String Theory,''
Mod.\ Phys.\ Lett.\ A {\bf 5}, 863 (1990).}
\lref\karch{A.~Clark, A.~Karch, P.~Kovtun and D.~Yamada,
``Construction of bosonic string theory on infinitely curved anti-de  Sitter space,''
arXiv:hep-th/0304107; A.~Karch,
arXiv:hep-th/0212041.}
\lref\lmrs{S.~M.~Lee, S.~Minwalla, M.~Rangamani and N.~Seiberg,
``Three-point functions of chiral operators in D = 4, N = 4 SYM at  large N,''
Adv.\ Theor.\ Math.\ Phys.\  {\bf 2}, 697 (1998)
[arXiv:hep-th/9806074].}
\lref\mettse{R.~R.~Metsaev and A.~A.~Tseytlin,
``Type IIB superstring action in AdS(5) x S(5) background,''
Nucl.\ Phys.\ B {\bf 533}, 109 (1998)
[arXiv:hep-th/9805028].}
\lref\berk{N.~Berkovits,
``Super-Poincare covariant quantization of the superstring,''
JHEP {\bf 0004}, 018 (2000)
[arXiv:hep-th/0001035].}
\lref\polchi{J.~Polchinski, Nucl.\ Phys.\ B {\bf 331}, 123, (1989).}

\Title
{\vbox{\baselineskip12pt
\hbox{hep-th/0308184}}}
{\vbox{\centerline{From Free Fields to $AdS$}}}

\centerline{Rajesh Gopakumar\foot{gopakumr@mri.ernet.in}}

\centerline{\sl Harish-Chandra Research Institute, Chhatnag Rd.,}
\centerline{\sl Jhusi, Allahabad, India 211019.}
\medskip

\vskip 0.8cm

\centerline{\bf Abstract}
\medskip
\noindent
Free ${\cal N}=4$ Super Yang-Mills theory (in the large $N$ limit) is dual to an, as yet, 
intractable closed string theory on $AdS_5\times S^5$. 
We aim to implement open-closed string duality in this system 
and thereby recast the free field correlation functions as amplitudes in $AdS$.
The basic strategy is to implement this duality directly on planar
field theory correlation functions
in the worldline (or first quantised) formulation.
The worldline loops (remnants of the worldsheet holes) close to form tree diagrams. 
These tree diagrams are then to be manifested as tree amplitudes in $AdS$ by a change of variables
on the worldline moduli space (i.e. Schwinger parameter space). 
Restricting to twist two operators,   
we are able to carry through this program for two and three point functions. 
However, it appears that this strategy can be implemented for 
four and higher point functions as well. An analogy to electrical networks
is very useful in this regard.

\vskip 0.5cm
\Date{August 2003}
\listtoc
\writetoc

\newsec{Introduction}

Over the last few years we have grown used to the idea of large $N$
gauge theories 
having a dual description in terms of gravitational theories in higher
dimensions \malda\gkp\witads . However, we need to remind ourselves that 
getting used to an idea is not the same as understanding it. 
It is fair to say that we do not really understand why or how (some) 
field theories 
reorganise themselves into a higher dimensional gravitational description. 

Open-closed string duality, we believe, is the underlying mechanism that 
drives these dualities. But 
except in the context of topological string dualities \gv\ov , we do not
explicitly understand how the holes in an open string description close up to form
closed string worldsheets. It would clearly be important to understand the nuts and bolts
of this mechanism better if we hope to shed further light on the miracles of
large $N$ dualities. 
  
A good idea is to begin with the simplest examples. From the field theory point
of view, a free theory is as simple as it gets. In particular,  ${\cal N}=4$ Super 
Yang-Mills at zero 
coupling is believed to be dual to 
string theory on a highly curved $AdS$ space (zero radius in string units) \sund . It is a 
measure of our lack of understanding of large $N$ dualities, that we know so little even
in this seemingly tractable limit. Interesting attempts
to understand the closed string sigma model, in this limit, have not yet yielded fruit \dmw\karch\polch . 

Therefore, as an alternative strategy, 
we might try to start from the free field theory, which is completely under 
control, and try to reconstruct the closed string theory, using as our guide the underlying
open-closed string equivalence. 

\subsec{Open-Closed String Duality}

Let us take this oppurtunity to elaborate a bit 
on our viewpoint on the general open-closed string equivalence. 
The leading large $N$ 
field theory
correlation functions (planar diagrams with some number of loops)
arise from planar (no handles) open string diagrams with some number of vertex 
insertions 
on its boundaries. Viewing this diagram in the closed string channel corresponds to 
gluing up the holes (while keeping the vertex insertions at finite separation). This is then 
interpreted as a closed string diagram\foot{To visualise the geometry of the gluing, 
think of the open string surface as a 
rubber sheet pinned at the locations of the vertex insertions. We can then imagine bringing 
together the boundaries of the rubber sheet (keeping the locations of the pins intact) 
and gluing them so as to obtain a genus zero surface with punctures at the locations 
of the pins.} with the same number of 
closed string vertex insertions, but with the gluing process having modified the background.
We will actually 
make a stronger working assumption which seems to be indicated by our analysis. 
We assume that this
open-closed string equivalence operates at the level of the worldsheet moduli space. 
An open string surface with particular locations of insertions and shape gets
associated with a particular closed string surface\foot{Of course, the usual 
counting of moduli
for the open and closed string give different dimensions. In our case, the matching
of moduli does not appear to be straightforward especially 
since, as we will see,  some of the open string moduli turn 
into parameters of the additional dimension. Presumably, this complication 
is related to the fact that
we do not have a CFT description for the closed string on $AdS$. It is important 
to understand this better.}. In other words, the 
gluing of the open string into the closed string is to be implemented on the {\it integrand} 
in moduli space. A change of variables on the moduli space would then exhibit this to be 
a closed string amplitude. 

Actually, this is probably too general a picture to be usefully implemented. What we will 
exploit is the simplification coming from the fact that we are working in the 
field theory limit. 
Since the ${\cal N}=4$ Yang-Mills theory is obtained as an $\apm\R 0$
limit of open string theory, one should really view the planar open string worldsheets
as reducing to planar worldlines. There is a precise sense in which this happens. The worldsheet
moduli space integral reduces to a worldline moduli space integral, 
more familiar as a Schwinger (or Feynman) parameter space integral for the Feynman 
diagrams (see, for instance, \berkos\divec\fmr ). The integrand reduces to 
a correlation function of worldline vertex operators in a first quantised formalism \polbk\strass . 
The particular simplification of the free field theory limit 
is that we have a Gaussian free particle action in this first quantised language.

So what we will aim to implement is the process of gluing up of planar worldline loops. It may 
seem puzzling at first that we have any kind of open-closed equivalence when the worldsheets
degenerate to lines. In fact, what we will see for the simplest class of field theory
correlation functions, is that the glued up version of the worldline 
is also 
a degenerate genus zero surface -- namely, with the topology of a tree.  The change of
variables on the moduli space, mentioned above, is the one natural to the description
of the tree. The new variables will have the interpretation as Schwinger parameters 
for the tree amplitude in $AdS$. Thus, even though the theory might be expected to be very stringy 
we find particle-like amplitudes at least for a class of states in $AdS$. 
It appears that the contributions for these amplitudes seem to come
from degenerate Riemann surfaces. We will comment on this further on. 
We will also make some
speculations in the concluding section on the appearance of ``fat'' closed string surfaces
from the glued up worldline. 

We could go ahead now and examine arbitrary correlators in the free theory in the 
worldline formalism. But free Yang-Mills (or super Yang-Mills),
in the large $N$ limit, has an exponentially large number of 
single-trace gauge invariant operators for a given dimension \sundb \pol . This is a reflection of  
its stringinesss. Implementing our strategy on arbitrary correlation 
functions of these operators is challenging because the worldlines can 
have a very complicated topology. 
Therefore, as a first step, it helps to focus on a subclass of simple 
operators for which the open-closed duality will be easiest to carry out.

\subsec{Twist Two Operators}

A very natural choice is to consider operators 
which are 
bilinear in the fields of the theory (but with arbitrary number of derivatives) 
transforming as symmetric traceless tensors of arbitrary spin. 
These twist two operators have several nice features.
One is that these operators form a 
set of higher spin conserved currents of the free theory. 
Another important feature is that they close 
amongst themselves under the OPE of the free theory\foot{Double trace operators like $(\Tr\Phi^2)^2$ 
are also present, but they correspond to multiparticle states of this sector.}
suggesting
some kind of consistent truncation to this subsector \mikh . 
Moreover, the leading order in $N$, connected  
$n$-point functions of these operators are particularly simple in the free theory, 
being given by a one loop diagram. In an open string  
(or equivalently, double line) representation
these are annulus diagrams with some number of insertions of gauge invariant 
operators (see Fig. 1). 
\fig{One loop open string diagram glued up into a closed string diagram.}
{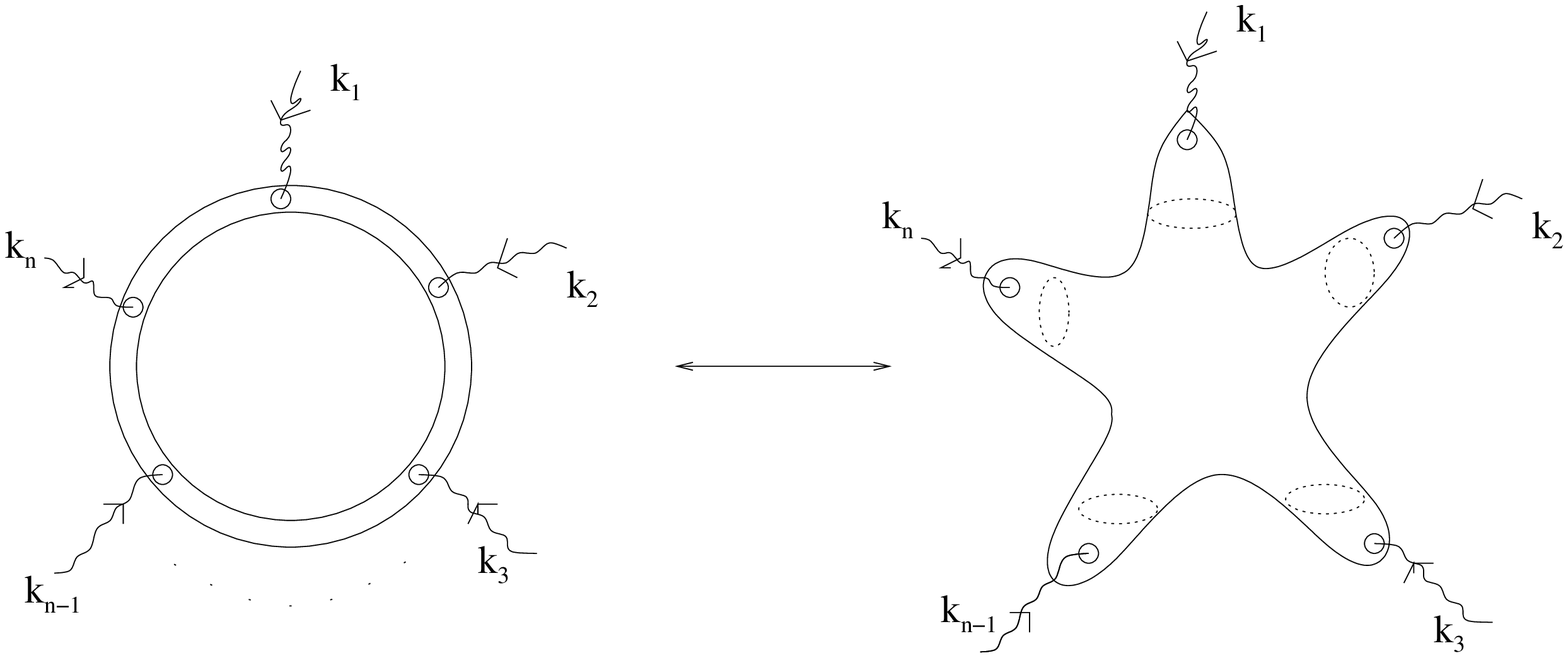}{6.0truein}
Thus topologically these are the simplest diagrams 
where we have just two holes (the inner and outer boundary) to close. In the field 
theory limit, the worldsheet reduces to a circular worldline (with
some number of insertions). So we just have to glue together a single worldline loop to
obtain trees. This simplifies our technical task.

The twist two operators are also naturally singled out from the 
closed string point of view. These 
operators correspond to the leading Regge trajectory of stringy excitations.
In the zero radius limit these are massless higher spin states in $AdS$, corresponding to the 
fact that the currents were conserved in the free theory \sundb\witt .  
In fact, classical interacting field theories of exactly these massless 
higher spin particles in $AdS_5$ have been studied by Vasiliev and others (see \vasrev\vas\sezsun\
for instance). 
That such classical theories exist at all, is some indication that perhaps there is a consistent 
truncation of the full string theory on $AdS_5\times S^5$  to this massless Regge 
trajectory \mikh . This also goes with the previous observation of the closure of the OPE in 
this sector of
the gauge theory.
It is also perhaps the explanation for why we find a description in terms of particle amplitudes upon
implementing open-closed duality on the twist two operators.
We should add that, even if true, this kind of consistent truncation 
would probably only hold for classical string theory (i.e. in the large $N$ limit). 
In any case, all these facts taken together 
suggest this sector of the theory to be a natural starting point 
for implementing our strategy. 

Klebanov and Polyakov \klepol\ have, in fact, attempted to isolate the dynamics of this sector 
by pointing out 
that the ``single-trace'' singlet operators in the $O(N)$ vector model are all bilinears
which have similar features to the gauge theory bilinears above and can thus be placed in 
exact correspondence with the massless higher spin states mentioned above. They therefore
conjectured that the large $N$ limit of the vector model was exactly dual to the 
classical Vasiliev theory\foot{Actually, their conjecture was that the $O(N)$ 
model in 3 dimensions, at its interacting (IR) 
fixed point is dual to the Vasiliev theory on 
$AdS_4$ (see \ruhl\por\peta\rajan\sumit\nemani\leigh\sezsuna\ for further work). 
The free field theory, or the UV fixed point was also 
conjectured to be dual to the Vasiliev theory on $AdS_4$, but with an 
inequivalent quantisation of the spin-0 field, something that is possible 
in $d=3$ \klepol .}. Many of our statements, therefore, can be carried over to the 
vector model at its UV fixed point. 
But as mentioned earlier, our strategy should enable us to go beyond the bilinears
once we have understood sufficently well the mechanism of the open-closed equivalence 
in this leading Regge trajectory case. We leave this for the future.

\subsec{The Organisation of the Paper}

Let us now chart out the flow of the paper.
In the next section we review the first quantised or worldline formalism. For reasons 
mentioned above, we will  
concentrate on the case where the worldline has the topology of a circle. The 
expressions for a general $n$-point function on the circle are well known and very similar to those
in string theory. As mentioned above, they take the form of an integral over 
``moduli space'' with an integrand which is the result of evaluating the correlation
function of vertex operators. 
Section 3 specialises to the case of three point functions and shows how the integrand
can be viewed as the circle glued into a three pronged tree. A similar thing happens 
trivially for the two point function as well. 
Section 4 connects this tree structure, that emerges from the worldline, to tree amplitudes
in $AdS$. To that end we first recast the usual bulk to boundary 
propagators in $AdS_{d+1}$ in a Schwinger 
representation. The key property that we want to exhibit is the close relationship to
the $d$-dimensional heat kernel. 
Using this representation, the three point function
in $AdS$ is seen to be simply related to the three pronged tree
amplitude of Sec.3 through a change of variables between the two Schwinger parametrisations.
This change of variables is {\it independent} of the external states and momenta. 
One interesting feature is that it is essentially the overall proper time modulus
that plays the role of
the additional dimension in $AdS_{d+1}$. 

Before proceeding further, we devote
Section 5 to an 
old analogy between the Schwinger parametrisation of Feynman diagrams and electrical networks.
Roughly speaking, for every Feynman graph 
the Schwinger parameters play the role of resistors while the external 
momenta play the role of currents. We exploit this analogy to understand 
how and why the loop got glued into a tree for the case of the one loop three point function.   
It also
provides intuition as to why we should expect a 
generalisation of this process to higher point functions. In fact, these considerations are not 
special to one loop correlators. Using  
the expressions for an arbitrary Feynman diagram in Schwinger parametrisation, one
might hope to implement the open-closed duality for arbitrary correlators. The intuition behind
the gluing of loops into trees is much the same. The expressions themselves are also
suggestive in their treelike structure. However, we do not pursue this at the moment.  
In Section 6 we look at the four point function. Guided by the electrical analogy,
we describe the equivalent tree diagram. This tree structure 
turns out to have the right form to be the 
four point function in $AdS$ with a sum over the different channels, including as  
intermediate states all the particles in the leading Regge trajectory. We leave the detailed
verification of this for the future.
Section 7 concludes with a summary, unfinished tasks and speculations. 
Appendix A deals with the two point function. Appendix B gives a convenient 
Schwinger representation of the scalar bulk-to-bulk propagator in $AdS$.
Appendix C touches on the relation 
to the heat kernel expansion and the UV/IR connection. 

Finally, a word about our formulas. In order to focus attention on the key physics aspects of 
various expressions, we have avoided cluttering them with overall
factors which are irrelevant to the considerations. 
Thus it would do well to remember that
the equality signs in equations are upto various such factors!

\newsec{The Worldline Formalism}

Let us consider a Euclidean free field theory in arbitrary
dimension $d$. (Though, we will mostly have in mind $d=4$ for application to $\CN=4$ Yang-Mills.) 
For concreteness, take a real scalar field $\Phi$
in the adjoint of $SU(N)$. This will suffice to illustrate the basic procedure. We will 
remark later on the generalisation to the full free super Yang-Mills. 

Let us consider the connected contribution to an $n$-point function of the gauge invariant 
operator $\Tr\Phi^2$
\eqn\nptdef{\G(x_1, x_2,\ldots x_n)={1\over N^2}\la \Tr\P^2(x_1)\Tr\P^2(x_2)
\ldots \Tr\P^2(x_n) \ra_{conn}.}
The corresponding momentum space correlator 
$\G(k_1, k_2,\ldots k_n)$ is given by the 
double line Feynman diagram in Fig.1 together with all other
possible permutations (of the indices $1,2\ldots n$) in the external leg insertions. 
One of the nice things about the worldline formalism is that all these diagrams are 
captured by a single worldline diagram of a circle with $n$ insertions where 
the location of each insertion is independent of the others leading to all 
possible orderings. 
\fig{One loop worldline diagram with $n$ insertions.}
{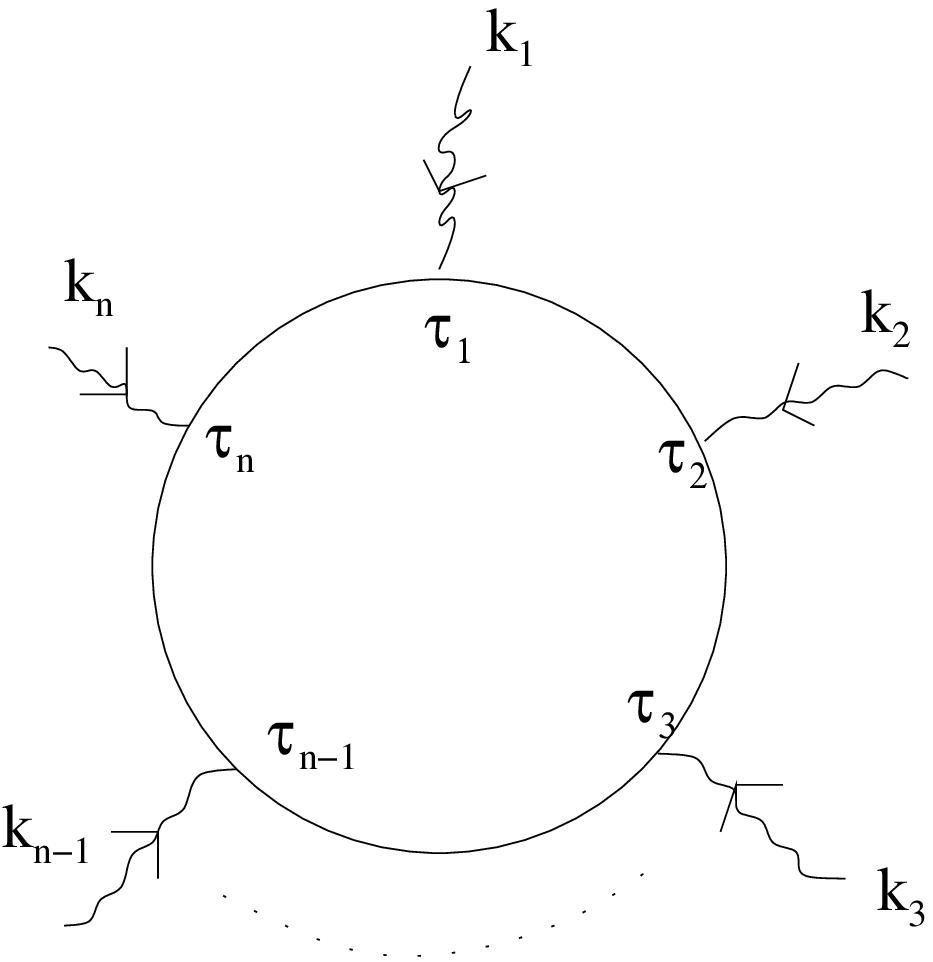}{3.0truein}
The locations (or proper times)
$\t_i$ of the insertions are worldline moduli. There is, in addition,
an overall modulus, the proper time $\t$ associated
with the invariant length of the circle. From the perspective of the first quantised action
of a free particle, 
$\t$ is the remnant of the world line diffeomorphism and plays a similar role 
to that of the conformal or Liouville mode in a worldsheet action (for a recent discussion, see 
\ssen .).

Then, the field theory correlation function takes a form very analogous to that 
of a string amplitude \polbk\strass .
\eqn\wlnpt{\eqalign{\G(k_1, k_2,\ldots k_n)=&\int [d\CM ]
\la e^{ik_1\cdot X(\t_1)}e^{ik_2\cdot X(\t_2)}\ldots e^{ik_n\cdot X(\t_n)}\ra \cr
\equiv &\int_0^{\infty}{d\t\over \t}\int_0^{\t}\prod_{i=1}^n d\t_i
\la e^{ik_1\cdot X(\t_1)}e^{ik_2\cdot X(\t_2)}\ldots e^{ik_n\cdot X(\t_n)}\ra .}}
The measure for $\t$ is consistent with the $U(1)$ invariance along the circle.
The correlator of the worldline vertex operators is evaluated with respect to 
the free particle action.
\eqn\wlcorr{\la \ldots\ra =\int [DX^{\mu}]\ldots\exp{(-{1\over 4}\int_0^{\t}dt(\p_tX)^2)}.}

A correlation function involving more general twist two scalar operators is again given by a 
one loop diagram as in Fig.2. However, the additional derivatives at each vertex are
reflected in the fact that the corresponding vertex operators will be linear combinations
(depending on the source momentum $k_i$) of 
$\p_tX^{\mu_1}\ldots\p_tX^{\mu_s}(\t_i)e^{ik_i\cdot X(\t_i)}$
\foot{For a given symmetric traceless tensor, one way to obtain the corresponding
vertex operator is to add a source term ($\propto e^{ik_i\cdot x}$) to the free field lagrangian, 
coupling to this operator. Integrating out the fields in this, still quadratic, lagrangian 
leads to a determinant which can be written as a proper time hamiltonian. 
In obtaining the determinant one integrates by parts the derivatives, which act on the 
$e^{ik_i\cdot x}$ and give factors of the momentum $k_i$. The proper time hamiltonian now
involves higher power of the derivatives, i.e. proper time momenta $p^{\mu}$, as well as the 
factor of $e^{ik_i\cdot x}$, in a definite ordering.  
The first quantised description involves going to 
the lagrangian description. To first order in the source, the additional term
in the free particle lagrangian is given by replacing the 
$p^{\mu}$, by $\p_tX^{\mu}$. This
term, linear in the source, is then the vertex operator corresponding to our original 
symmetric traceless tensor.}. Unlike the infinitely many oscillators of the string, 
here the point particle
equation of motion $\p_t^2X^{\mu}=0$, leads to a restricted class of vertex 
operators -- those of a single Regge trajectory. 

In evaluating an arbitrary correlator of bilinears, it is only the vertex operators that 
are modified as above. 
The other aspects of the worldline expression remain the same.
In particular, we have the same integral over worldline moduli. Though the details of 
the tensor structure of the more general vertex operators will be important for detailed
matching with amplitudes in $AdS$, the primary feature of gluing up of the loop will arise
from the worldline correlators of $e^{ik_i\cdot X(\t_i)}$. 
Therefore we will mostly
concentrate on the $n$-point function in Eq.\wlnpt .

The correlation function of vertex operators appearing in Eq.\wlnpt\ is easily evaluated
by  performing the Gaussian integral of Eq. \wlcorr . The result is that the 
integrand in moduli space takes the explicit form (see \polbk\ for instance)
\eqn\modint{
\la e^{ik_1\cdot X(\t_1)}e^{ik_2\cdot X(\t_2)}\ldots e^{ik_n\cdot X(\t_n)}\ra
=\d^{(d)}(\sum_ik_i){1\over \t^{d\over 2}}\exp{(-\half\sum_{i\neq j}^nk_i\cdot k_jG(\t_i,\t_j))},}
where   
\eqn\green{G(\t_i,\t_j))=-{\t_{ij}(\t-\t_{ij})\over \t};~~~~~~~~~(\t_{ij}=|\t_i-\t_j|),}
is the appropriate Green's function on the circle. The $\delta$ function enforcing momentum 
conservation comes from the zero mode integral over the $X$'s. The factor of ${1\over \t^{d\over 2}}$
is from the determinant of the non-zero modes.

To evaluate the correlation function of the more general vertex operators, mentioned above,
is equally straightforward, involving just keeping track of more indices. As usual, 
an efficient 
way to obtain them is to introduce a source term $j\cdot\p_tX$ in the worldline action and 
carry out the Gaussian integral and take the required number of functional derivatives.
All that essentially changes in Eq.\modint\ is that one has polynomial
factors in the external momenta, together with time derivatives of the Green's function,
multiplying the Gaussian factor in \modint .

Going back to Eq.\wlnpt , the complete expression for the $n$-point function is  
\eqn\wlint{
\G(k_1, k_2,\ldots k_n)=\d^{(d)}(\sum_ik_i)
\int_0^{\infty}{d\t\over \t^{{d\over 2}+1}}\int_0^{\t}\prod_{i=1}^n d\t_i
\exp{(-\half\sum_{i\neq j}^nk_i\cdot k_jG(\t_i,\t_j))}.}
It is this expression that will be our main focus of attention. We will examine it more 
closely for the three and (to a lesser extent) the four point function. 
In these cases, we will see how
this integral over moduli space for a loop can
be viewed in terms of contributions from tree graphs.  

Finally, a word about further generalisations. Firstly, when there are several species
of scalar fields, then the only Wick contractions that survive are ones where the flavour
indices contract. This means that only a subset of permutations of the external legs give
a nonzero answer. This effectively translates into a truncation of the regime of 
integration of the moduli $\t_i$. This issue only arises for four and higher point 
functions.

Fermionic and gauge bilinears can also be incorporated into the worldline formalism \bdh .
This is best done 
by including a worldline grassmann superpartner $\psi^{\mu}$
to the $X^{\mu}$ and appropriately
supersymmetrising the free worldline action. The natural description for the action 
and vertex operators is in terms of a worldline superfield which is integrated over a 
supermoduli space. The expressions for the one loop correlation functions
are reductions of analogous superstring ones. Once again the essential features are captured by 
Eq.\wlint .
We refer the reader to \strass\ and the review \schb\
for more details and references. 

\newsec{The Three Point Function}

Having set up the worldline formalism for the $n$-point function of bilinears, we will 
now see how the worldline
gluing process takes us from the one loop diagram to a treelike structure,
in the particular 
case of the three point function. Logically, one should start with  
the two point function. But 
the gluing process is somewhat trivial there and we shall relegate it's discussion
to the appendix. 

Once again, to keep things uncluttered we will
begin with the $n=3$ case of Eq. \nptdef , or rather its momentum space version in \wlnpt ,
which was evaluated in \wlint .
\eqn\thrpt{\G(k_1, k_2, k_3)=\d^{(d)}(\sum_ik_i)
\int_0^{\infty}{d\t\over \t^{{d\over 2}+1}}\int_0^{\t}\prod_{i=1}^3 d\t_i
\exp{(-\half\sum_{i\neq j}^3k_i\cdot k_jG(\t_i,\t_j))}.}

The momentum conserving delta function  
helps in simplifying the kinematic invariants appearing in the exponent, in
Eq.\thrpt . Thus $2k_1\cdot k_2=k_3^2-k_1^2-k_2^2$ etc. Making the change of variables
$\t_{12}=\t\a_3, \t_{23}=\t\a_1, \t_{31}=\t\a_2$ (with $\sum_i\a_i=1$) yields the 
simpler form
\eqn\thrept{\eqalign{\G(k_1, k_2, k_3)= &\d^{(d)}(\sum_ik_i)
\int_0^{\infty}{d\t\over \t^{{d\over 2}+1}}\t^3\int_0^{1}\prod_{i=1}^3 
d\a_i\d(\sum_i\a_i-1)\cr
\times &\exp{\{-\t(k_1^2\a_2\a_3+k_2^2\a_3\a_1+k_3^2\a_1\a_2)\}}.}}
We recognise the integrand to be 
\eqn\momglue{\exp{\{-\sum_i\t k_i^2\a_j\a_k\}}
=\prod_{i=1}^3\la k_i|\exp{[\t\a_j\a_k\lform]}|k_i\ra ,}
where the indices $i,j,k$ are a cyclic permutation of $\{1,2,3\}$
and $\lform$ is the $d$-dimensional Laplacian whose eigenkets are denoted by $|k\ra$.

That this heat kernel representation
is already a glued up version of the original loop diagram can be most clearly 
exhibited by going to position space. We can easily do the Gaussian integral over momentum
(after introducing a centre of mass 
variable $z$ as lagrange multiplier for the delta function).
\eqn\posglue{\eqalign{\G(x_1,x_2,x_3)=&\int\prod_{i=1}^3d^dk_ie^{-ik_i\cdot x_i}\G(k_1,k_2,k_3)\cr
=&\int_0^{\infty}{d\t\over \t^{{d\over 2}+1}}\t^3\int_0^{1}\prod_{i=1}^3 
d\a_i\d(\sum_i\a_i-1){1\over \t^{{3d\over 2}}(\prod\a_i)^d}\int d^dz
\exp{\{-{1\over 4}\sum_{i=1}^3{(x_i-z)^2\over \t\a_j\a_k}\}}\cr
=&\int_0^{\infty}{d\t\over \t^{{d\over 2}+1}}\t^3\int_0^{1}\prod_{i=1}^3 
d\a_i\d(\sum_i\a_i-1)\int d^dz\prod_{i=1}^3\la x_i|\exp{[\t\a_j\a_k\lform]}|z\ra .}} 
The last line exhibits the position space heat kernel representation clearly as a 
tree amplitude -- the product of free particle amplitudes to propagate from each $x_i$ to
a common vertex $z$ (which is then integrated over).\foot{That the one loop three point function
can be viewed as a tree in this way was noticed by \bv .}  
Recall that the heat kernel or propagator in position space is
given by
\eqn\prop{\la x|e^{t\lform}|y\ra={1\over (4\pi t)^{d\over 2}}e^{-{(x-y)^2\over 4t}}.}
Pictorially, we may depict the process of gluing as in Fig. 3.
\fig{One loop three point function glued up into a three pronged tree. The relation between 
the Schwinger parameters of the sides of the loop and the legs of the tree are also shown.}
{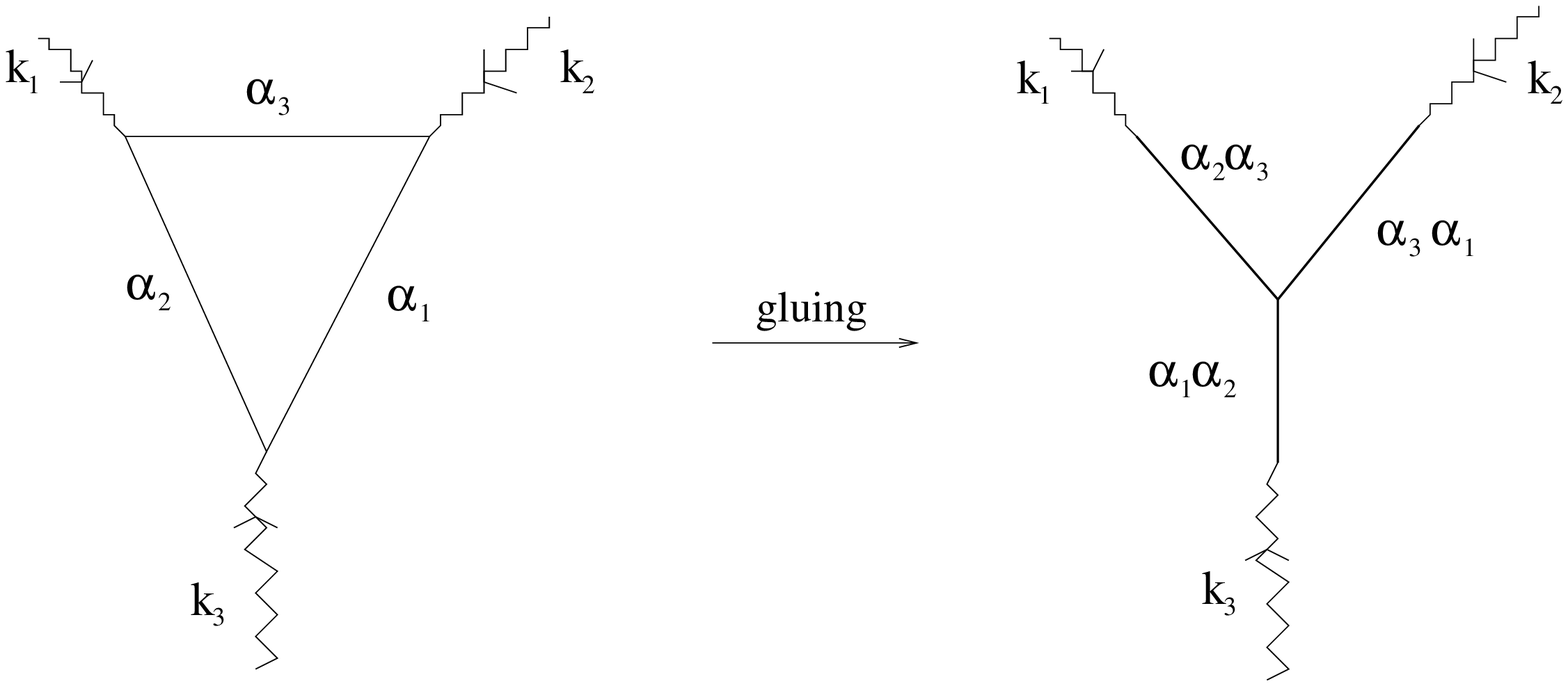}{6.0truein}

In the next section we will see that this tree is precisely a tree amplitude in $AdS_{d+1}$
once we make a change of variables in Eq.\posglue\ into Schwinger parameters for the tree. 
But before we proceed, let us comment on the three point correlation 
function of arbitrary bilinears. 
As mentioned in the last section, the only changes are multiplicative factors 
consisting of polynomials in the external
momenta (and $\a_i$). Since the crucial Gaussian factor is unchanged, we see that 
in position space, we continue to have the tree structure of 
$\prod_{i=1}^3\la x_i|\exp{[\t\a_j\a_k\lform]}|z\ra$. The additional terms are 
multiplicative polynomials in the $(x_i-z)^{\mu}$. Similar, slightly generalised remarks apply 
to the case of fermionic and gauge bilinears. 
Thus the property of gluing depicted in Fig.3  
is universal to all the one loop correlation functions. We will better understand the 
underlying reason  for this in Sec. 5.

\newsec{The Three Point Tree Amplitude in $AdS$}

\subsec{The Bulk to Boundary Propagator in $AdS$}

To compare the tree of the previous section with the tree amplitudes in $AdS$ we 
will find it useful to write the bulk to boundary propagators for various fields in 
a somewhat unconventional manner. We have mostly focussed attention on correlation functions
of $\Tr\Phi^2$.
The dimension of this operator is $(d-2)$ in the $d$-dimensional free theory. 
The $AdS/CFT$ dictionary \gkp\witads\ then tells us that it couples to a scalar field in $AdS$ with 
$m^2=-2(d-2)$. 
Hence
let us start with the bulk to boundary propagator of such a scalar field. 
This is a solution of the 
wave equation for a scalar in $AdS_{d+1}$ (where we set the radius of $AdS$ to one) 
\eqn\adswv{[-z_0^2\p^2_{z_0}+(d-1)z_0\p_{z_0}-z_0^2 \lform -2(d-2)]K=0,}
which is proportional to a delta function when the 
position in the bulk approaches the boundary. Here we are working with Euclidean $AdS_{d+1}$
in the natural Poincare coordinates
\eqn\adsmet{ds^2={d\vec{z}^2+dz_0^2\over z_0^2}}
and $\lform$ is the $d$-dimensional Laplacian in the directions $\vec{z}$ 
(or simply $z$ when there is no risk of confusion) that parametrise the 
boundary on which the dual field theory resides.

In terms of the coordinate $t=z_0^2$ and $K=t\tilde{K}$, Eq.\adswv\ takes the form
\eqn\adseq{[2(d-6){\p\tilde{K}\over \p t}- \lform \tilde{K}-4t{\p^2\tilde{K}\over \p t^2}]=0.}
If the last term were absent, this would have been the heat equation and the solution would 
have simply been the heat kernel $e^{t\lform}$. However, solutions to \adseq\ can also be
expressed in terms of the heat kernel. Thus the usual bulk to boundary 
propagator is given by a solution of the form 
\eqn\htk{\tilde{K}(t)=\int_0^{\infty}d\r\r^{{d\over 2}-3}e^{-\r}e^{{t\over 4\r}\lform}.}
Thus, $K(t)=t\tilde{K}(t)$ is expressed in terms of a convolution over a heat kernel
in terms of the parameter $\r$. It is easy to verify that this is just a Schwinger 
parametrisation
of the familiar bulk to boundary propagator (with $\Delta=d-2$)\witads\ 
\eqn\usualk{K(x, z; z_0=t^{\half})=
\big({t^{\half}\over t+(x-z)^2}\big)^{d-2}=\la x|K(t)|z \ra .}
In momentum space \gkp , Eq.\htk\ is nothing more than an integral representation of the bessel 
function that the bulk to boundary propagator is proportional to.

This close relation to the $d$-dimensional heat kernel 
is the main reason why the glued up tree of the previous section can be related to a tree
amplitude in $AdS$. Though we have shown this for the scalar and that too for its bulk to
boundary propagator, it is clear that wave equations for higher spin particles in $AdS$
can also be put into a similar form as \adseq\ which exhibits the close relation to 
the heat kernel. Similarly bulk to bulk propagators will also be expressed in terms of 
$d$-dimensional heat kernels, as we will see explicitly in Sec. 6.

\subsec{The Three Point Function}

Let us write down the tree amplitude for the three point function of the above scalar in this Schwinger
representation. If there are no derivatives in the cubic couplings of scalars (something that
can presumably be achieved by a field redefinition \lmrs ) the point particle amplitude is
(using Eqs. \usualk , \htk\ )
\eqn\threetree{\eqalign{\G(x_1, x_2, x_3)
= &\int d^dz\int_0^{\infty}{dz_0\over z_0^{d+1}}
\prod_{i=1}^3 K(x_i, z; z_0)\cr
= &\int_0^{\infty}{dt\over t^{{d\over 2}+1}}t^3\int_0^{\infty}\prod_{i=1}^3
d\r_i\r_i^{{d\over 2}-3}e^{-\r_i}\int d^dz\prod_{i=1}^3\la x_i|e^{{t\over 4\r_i}\lform}|z\ra.}}
Note the close similarity with the
integrand of the
worldline expression Eq.\posglue\ particularly the striking closeness between the
radial coordinate $t$ and the proper time $\t$. However, here instead of integrating over 
the worldline moduli of the loop
we have an integral over Schwinger parameters for the tree. There is a simple
change of variables 
between the two which makes the two integrals identical. This is suggested by the relations in 
Fig.3. between the loop and the tree. We simply have to put $\r_i=\r\a_i$, where $\r=\sum_i\r_i$.
Which we can implement by introducing $\int_0^{\infty}\d\r\d(\r-\sum_i\r_i)=1$ into the integral 
and changing to variables $\a_i$. Finally, we make the change
\eqn\proptime{t=4\t\r(\prod_{i=1}^3\a_i),}
which relates the proper time $\t$ to the $AdS$ radial coordinate $t$. The integral over $\r$ 
decouples, only contributing to the overall constant which we have dropped all along,
and  Eq.\threetree\ becomes Eq.\posglue .
   
A number of comments are in order here. The change of variables that we made was independent 
of the external momenta or positions and even of the number of space time dimensions. It is the 
kind of change of variables one might expect in going between a parametrisation of open string
moduli space and one of the closed string. 
In fact, in 
generalising to the three point function of arbitrary bilinears, we expect 
that the same change of variables will be sufficent. This is essentially because 
both the exponent and the measure on moduli space continue to be the
same for the general three point function. Of course, it is not guaranteed that the 
multiplicative tensor structures will work out right. 

This is where, we believe, the supersymmetry and the special field content of $d=4, \CN=4$ 
Yang-Mills will play a special role. After all what we have done so far works for any free 
scalar theory in any dimension. 
It is likely that it is only in the case of $\CN=4$ Yang-Mills that  
the tensor structure encoded in the multiplicative factors would also match with that
from the bulk to boundary propagators for massless higher spin paricles.

There is some evidence for this contention. 
Indeed, in the early days of the $AdS$/CFT correspondence, 
a free field computation of the (two and) three point function of R-currents (which are
bilinears in the fields)
in $\CN=4$   
Yang-Mills was compared to supergravity \cnss . The authors of \cnss\
employed a Schwinger parametrisation of 
the one loop diagram and matched the resulting 
integral expression with that of the supergravity integral, again in 
a parametrised representation. Making a change of variables, essentially equivalent to
that above, they found an explicit agreement of the tensor structures\foot{We would like to thank 
K. Schalm for drawing our attention to \cnss .}. It was important 
for them that the contribution from both scalars and fermions to the R-current correlators 
were taken into account to get the exact matching. We take this as evidence that 
the special properties of $\CN=4$ Yang-Mills
are likely to play a role in ensuring a detailed matching of tensor structures. 
A related observation is regarding the conformal anomaly. (Similar considerations apply to the two 
and three point functions of the stress tensor. The detailed matching of these tensor structures
in the bulk and the boundary was carried out in \arutfrol .) 
The conformal anomaly in $d=4$, for example, is a linear combination of two independent curvature 
invariants. The particular combination depends on the field content of the theory.
An $AdS_5$ calculation, on the other hand,
gives a definite combination of these two invariants \hsken . 
One needs the full field content of $\CN=4$ Yang-Mills, to get this particular combination.
Thus the the bosonic vector model in four dimensions, for example,  
cannot possibly arise from an $AdS_5$ calculation since a different linear combination 
of the curvature invariants arises in the two computations\foot{We would like to thank 
I. Klebanov and K. Skenderis for discussions on this point.}. 

One of the very interesting features in making the connection between the worldline picture and 
amplitudes in $AdS$ is that the proper time on the worldline is more or less directly related to
the radial direction in $AdS$, as seen in Eq. \proptime\ or in the measures of \posglue\ and  
\threetree . This is not altogether unexpected. The propertime $\t$ is a measure
of the energy scale in the field theory and it has been seen in various circumstances
that the radial coordinate in $AdS$ plays a very similar role. For instance, a UV cutoff 
in the field theory can be implemented by cutting off the modulus integral at small $\t$. From
\proptime , this effectively translates into an IR cutoff in the radial coordinate $t$ in $AdS$.
This is also apparent from the fact that 
$\t$ is the remnant of the modulus of the open string annulus, and 
the small $\t$ regime is where the annulus captures the long 
distance (IR) propagation in the closed string channel. 
Another source of our intuition for why the 
proper time should play the role of the extra dimension comes from the observation that $\t$
represents the worldline conformal factor, and so in a loose sense, is a Liouville mode\foot{We
thank S. Wadia for pointing this out.}. Therefore
it fits in with the idea of the Liouville mode being 
the origin of the extra dimension in the $AdS/CFT$ correspondence \polb\polc .
(Note that the idea of the Liouville direction playing the role of an additional spacetime dimension
in noncritical string theory goes back to \satchi\spenta\polchi .)  
It will be very interesting to flesh this connection out further.

\newsec{The Electrical Network Analogy}

At this stage, 
it might seem that the results of the previous section are due to the special nature of 
(two and) three point functions which are largely constrained by conformal invariance\foot{For 
correlators of higher spin operators there are a finite number of tensor structures 
consistent with conformal invariance. The relative coefficents are undetermined.}.  
What we would like to motivate in this section is that the basic mechanism which is operating 
is indeed open-closed duality, (in the limit where the Riemann surfaces on both sides 
have degenerated to graphs). And that this mechanism can generalise to arbitrary diagrams. 

We recall that there were two main 
steps in the process of going from the field theory 3-point function
to the $AdS$ amplitude. The first was to argue that the worldline formulation of the
field theory loop could be seen in terms of trees involving free particle heat kernels. 
The second was to show that these trees were indeed tree diagrams in $AdS$. The latter turned out 
to be true essentially because
the wave equation in $AdS$ implied 
a close connection between propagators in $AdS_{d+1}$ and free particle heat kernels 
in $d$ dimensions. A change of variables on the moduli then demonstrated the identity of the 
two tree amplitudes. The first step of gluing loops into trees
is the one where the geometric mechanism of open-closed duality seems to be operating.  
To better understand how this operates, and generalises to arbitrary correlation functions, 
it will be very useful to revive an old analogy between Feynman diagrams and electrical networks. 

The first indication that such an analogy might be present, and important for us,
is the observation that the loop-tree duality in Fig. 3 is  
similar to the standard 
``star-delta'' equivalence in electrical networks\foot{We thank 
Justin David for this observation which
was instrumental in our pursuing this line of thought.}. In fact, 
there is a precise connection. If one views the ``delta'' (loop) diagram in Fig.3 as an 
electrical network with resistances $R_i$ in each of the sides, then this network is exactly
equivalent to that of the ``star'' or tree diagram with resistances $(R_jR_k/\sum R_i)$ 
on the legs as 
shown in the figure\foot{Star-triangle relations crop up very often in physics and mathematics. 
In a related context, see \petkou .}. 
This may be verified using elementary considerations of Kirchoff's laws.
The essential idea involves eliminating the current flowing in the loop from the equations
so that we are reduced to an equivalent tree diagram without that loop 
(see, for instance, \guil ). 

This is not just a coincidence. There is an analogy between Feynman diagrams and 
electrical networks going back to the 1960's (see, for instance, chapter 18 of \BD\ ). 
An arbitrary Feynman diagram, 
expressed in Schwinger (or Feynman) parametrisation,  
has a natural interpretation in electrical network terms. The Schwinger moduli
can be identified with resistances. And the external, as well as 
internal, momenta with currents flowing in the respective legs. 
The process of carrying out the integrals over loop momenta is then equivalent to elimination
of the internal currents using Kirchoff's laws\foot{This can be seen in the 
Schwinger parametrisation where one is performing Gaussian integrals over the internal momenta. 
The exponent has the interpretation as being the power consumed in the Feynman diagram. 
The Gaussian saddle points are precisely the Kirchoff equations for voltages.}.
The result  is a generalisation 
of our one loop worldline expressions -- an integral over Schwinger moduli space of an integrand
that depends on the external momenta. In particular, the crucial piece is a 
Gaussian exponent as in \modint , which is proportional to the power 
consumed in the equivalent circuit. This is clearly 
visible in \thrept\ where the exponential factor is the power consumed in 
the equivalent tree circuit of Fig. 3. 
For a recent review of (and earlier references to) 
the expressions for a general Feynman diagram as well as their electrical interpretation, 
see \lam .   

What is of interest to us is the tree structure obtained after 
elimination of the loop momenta/currents. In the language of circuits it
is intuitively plausible that this process 
of elimination of loop currents results in an equivalent tree structure. The external
currents in various linear combinations would flow through the various legs of this tree.  
For instance, the expression for the power in the equivalent circuit, which appears as the 
Gaussian exponent in the integral can be written down explicitly. 
For a general $l$ loop diagram, it
is given in graph theoretic terms \lam 
\eqn\power{P(\a,k)=\D(\a)^{-1}\sum_{T_2}(\prod^{l+1}\a)(\sum k)^2.} 
Here, $\a_i$ are the Schwinger parameters for the various internal legs of the loop. 
the sum is over various 1-trees and 2-trees obtained from the original loop diagram. 
A 1-tree is obtained by cutting the $l$ loop diagram at $l$ lines so as to make a connected 
tree. While a 2-tree is obtained by cutting the loop at $l+1$ lines so as to form two 
disjoint trees. $\D(\a)$ is then given by a sum, over the set $T_1$ of all 1-trees, of the product
of the $\a_i$ of all the cut lines. In the case of a one loop diagram, this is simply $\sum \a_i$.
The sum over $T_2$ indicates a sum over the set of 
all two trees, where the product is over the $\a_i$ of the 
$l+1$ cut lines. And $(\sum k)$ is understood to be the sum over all those external momenta $k_i$ 
which flow into (either) one of the two trees. It is easy to verify, for example, that in the case of the 
three point function this tallies with the exponent in \thrept . 

This expression can be interpreted as the power dissipated in an equivalent tree circuit 
in which currents $(\sum k)$ flow in legs whose resistances are $\D(\a)^{-1}(\prod^{l+1}\a)$. 
The topology of this tree circuit seems somewhat intricate in general. We will examine the 
case of the one loop four point function somewhat more in the next section. However, 
we will postpone a more
general analysis to future work. But hopefully, what should be clear from the above considerations
is that the gluing up of loops into trees is not particular to two or three point functions.
What we are seeing is an implementation of open
closed string duality, in the limit of degenerate worldsheets. 
And the electrical analogy gives us useful intuition for visualising this process.

When there are multiple loops it is likely that the glued up tree is effectively a thick or 
``fat'' worldsheet, at least in the limit where a large number of loops are present. 
The latter would be true for correlation functions of, say, $\Tr \Phi^{J}$ for large $J$. 
It would be nice to make a connection with the BMN picture \bmn\ of a closed string worldsheet 
emerging from operators like this carrying a large number of bits.

\newsec{The Four Point Function}

Armed with the intuition from electrical networks, we will take a first look at the 
four point function. Here we will only try to convince the reader that the worldline diagram 
does glue up in the right way as expected from the duality to $AdS$. A detailed check will be 
postponed to the future. 

As usual we will restrict our consideration to the four point function of $\Tr\Phi^2$, only 
briefly indicating the generalisations.
The worldline expression is given from \wlnpt\ to be
\eqn\fourpt{\G(k_1, k_2, k_3, k_4)=\int_0^{\infty}{d\t\over \t}\int_0^{\t}\prod_{i=1}^4 d\t_i
\la e^{ik_1\cdot X(\t_1)}e^{ik_2\cdot X(\t_2)}e^{ik_3\cdot X(\t_4)}e^{ik_4\cdot X(\t_4)}\ra .}
The integral over moduli space can be broken into six
cyclically inequivalent orderings of the four insertions. Of these three are 
related to the others by a worldline reflection $\t \R-\t$. The three inequivalent orderings
correspond to three inequivalent
Feynman diagrams that can contribute to this amplitude. As mentioned in Sec. 2, when there
are several flavours of scalars (as in $\CN=4$ Super Yang-Mills) then some of these diagrams 
might be absent due to the structure of the flavour indices. This can easily be incorporated into
the considerations below.

Let us look at a specific time ordering $(1234)$ of the insertions around the circle. The 
worldline expression is given by Eq.\wlint\ with the caveat that the integral over $\t_i$
is restricted to the above time ordered domain. The all important Gaussian factor 
in the integrand can be rewritten using the delta function constraint on the momentum as  
\eqn\gaussfour{e^{-\half\sum_{i\neq j}^4k_i\cdot k_jG(\t_i,\t_j)}=
e^{-\t(\a_4\a_1k_1^2+\a_1\a_2k_2^2+ 
\a_2\a_3k_3^2+\a_3\a_4k_4^2+\a_2\a_4(k_1+k_2)^2+\a_1\a_3(k_1+k_4)^2)}.}
Here $\t_{41}=\t\a_4,\t_{12}=\t\a_1$ etc. The exponent on the right hand side
tallies with the general Schwinger 
parametrisation expression quoted in \power . As we saw in the previous section this exponent
has the interpretation as the electrical power dissipated in the equivalent circuit obtained 
after eliminating the loop current. 

What is the equivalent circuit in this case? It is as shown in the right of Fig.4\foot{This figure 
is actually adapted from a textbook on electrical networks 
(see page 136 of \guil )!}. Elementary circuit analysis 
shows that this is the equivalent circuit. One quick way to verify this is to 
see that the exponent in \gaussfour\ is proportional to 
the power dissipated in the thick lines on the right.
Note that the topology is now more complicated than in the case of the three point function. 
In fact, the set of equivalent resistors, shown with thick lines, is no longer fully connected.
The horizontal resistor (which has current $(k_1+k_4)$ flowing through it) is disconnected 
from the 
others. Of course, we have drawn the  equivalent circuit in the ``$s$-channel''. This is an 
arbitrary choice. One could equally
well have drawn it in the ``$t$-channel'', in which case the vertical line would have been 
the disconnected one. 
\fig{One loop four point function and the equivalent tree. 
In the language of the electrical network, the thick lines are the 
equivalent resistors of the tree. The dotted lines complete the rest of the circuit.
The whole diagram is best thought of as drawn on a sphere with the dotted lines going 
behind.}   
{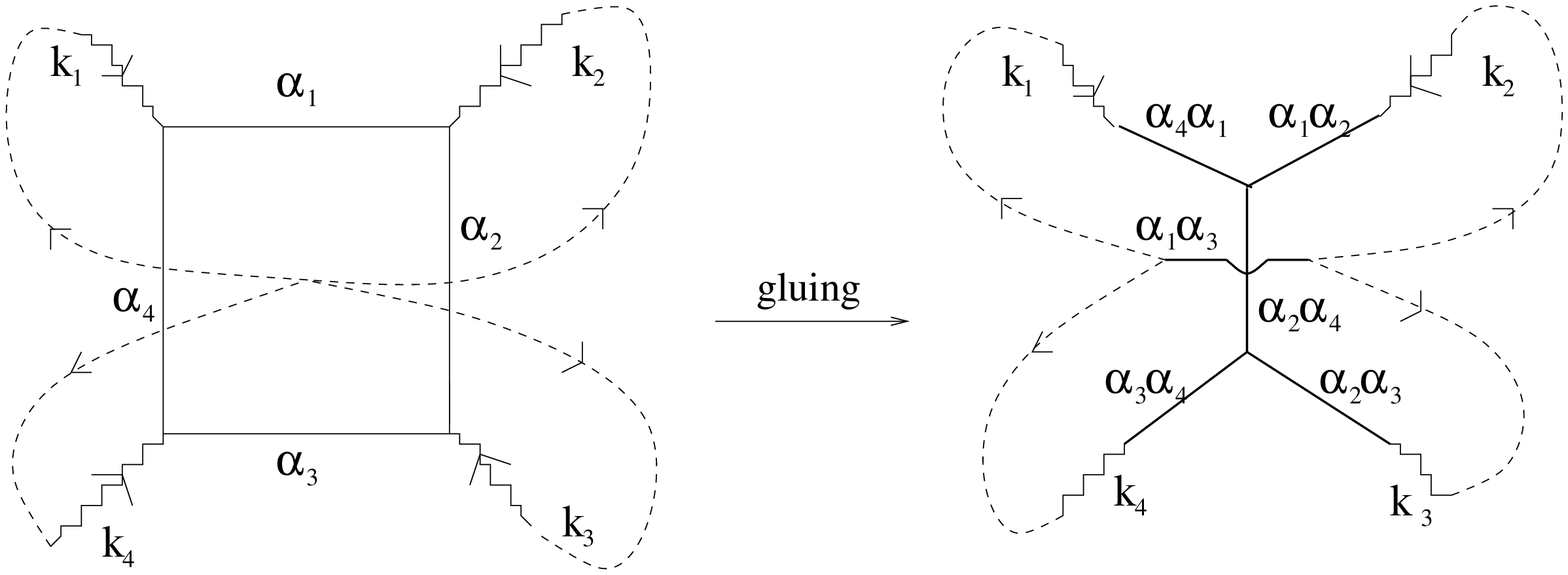}{6.0truein}

We would like to claim that
this tree structure, which the loop is glued into, is what is needed for the $AdS/CFT$
duality. If we consider the four point amplitude in $AdS$ in a point particle limit, 
it is given by a set of tree 
diagrams (see \dfmmr\ and references therein to the large literature
on $AdS$ four point functions. In particular, the four point functions of the lowest twist
two operators have been studied in detail, both perturbatively and at strong coupling in 
\glebfrpet ). 
In the point particle limit (unlike in a worldsheet description) one 
separately sums over diagrams in $s,t,u$ channels built from three point vertices.
These diagrams are as in the tree of Fig. 4 (minus the horizontal line). 
In a given channel, say $s$, there can be an infinite set of intermediate states. We 
will now try to argue that the horizontal line in Fig. 4 captures an infinite summation in the
$s$-channel. 

To interpret the horizontal resistor as a sum over infinitely many states, let us go back to
\gaussfour . If we expand the corresponding 
exponential piece $e^{-\t\a_1\a_3(k_1+k_4)^2}$, we get an 
infinite series in powers of $t=(k_1+k_4)^2$. A term $\sim t^J$
immediately suggests a contribution from 
a state of spin $J$ in the intermediate channel. 
But we could have equally well viewed this diagram in the $t$-channel, the $s-t$
symmetry would imply an infinite summation in this channel too.
There are two other inequivalent orderings of insertions, 
namely, $(1324)$ and $(1243)$ which have $t-u$ and $s-u$ symmetries respectively. The sum has complete
$s-t-u$ symmetry. 
We can
expand each of these inequivalent diagrams twice,\foot{Note that in our way of reckoning, each 
ordering appears twice, once in cyclic and again in anticyclic order. So we do not introduce any new 
factors of two in expanding each diagram in two ways.} once in each of the respective channels and 
sum over all these orderings. We then rearrange the final answer
as a sum over all channels (each channel getting infinite summation contributions from two different 
orderings). As mentioned above, the powers appearing in the infinite
sum are suggestive of higher spin 
intermediate states. 
Thus the worldline expression
{\it prima facie} has the 
right structure to assemble itself into 
an $AdS$ amplitude with different channels, each involving higher spin intermediate states.
We must emphasise again that it is because we are viewing the $AdS$ amplitude in a point particle 
language that we obtain a sum over different channels, each with an infinite number of intermediate 
states. This is more like a closed string field theory representation which pieces together different
regions of moduli space to achieve a dual answer.\foot{We thank L. Rastelli, A. Sen and E. Witten for
useful discussions in this regard.} 
In this context, we should mention that there could also be a four point 
contact term, in principle. A more careful examination of the amplitude will
be necessary to disentangle such a contribution.

Another related viewpoint also indicates an infinite tower of intermediate states in each channel. 
Since we have taken the ordering $(1234)$, the vertex operators $e^{ik_1\cdot X(\t_1)}$
and $e^{ik_2\cdot X(\t_2)}$ are adjacently inserted. 
Remembering that these vertex operators
are normal ordered, we can write the exact 
worldline operator product
\eqn\op{e^{ik_1\cdot X(\t_1)}e^{ik_2\cdot X(\t_2)}=e^{ik_1\cdot X(\t_1)+ik_2\cdot X(\t_2)}
e^{-k_1\cdot k_2G(\t_1 ,\t_2)}.}
If we expand the non-local vertex operator $e^{ik_1\cdot X(\t_1)+ik_2\cdot X(\t_2)}$
in powers of the separation $\t_{12}$, one obtains vertex operators of the form
$\p_tX^{\mu_1}\ldots\p_tX^{\mu_s}(\t_0)e^{i(k_1+k_2)\cdot X(\t_0)}$ where $\t_0$ is the midpoint
of $\t_1$ and $\t_2$. We have again used the equation of motion $\p_t^2X^{\mu}=0$. 
This indicates that as intermediate states in the $s$-channel we will have all the
higher spin particles of the leading Regge trajectory. In fact, this actually suggests that we 
do not have particles 
from other Regge trajectories appearing as intermediate states. This is 
in line with the comments in the introduction about the possibility of a 
consistent truncation to the leading Regge trajectory. 

For four point functions of more general twist two operators the logic is very similar
since we have the same Gaussian factor. There is, in addition, a multiplicative factor
in the momenta which contributes to the spin of the exchanged state. The intermediate states
are still in the leading Regge trajectory following the arguments of the previous paragraph. 

\subsec{The $AdS$ Four Point Function}

Though we will not make a detailed comparison with the four point function in $AdS$ at present,
we will look at the glued up worldline answer in position space and exhibit the closeness 
to $AdS$ amplitudes. Using \wlint\ and \gaussfour\ 
the full worldline expression for the four point function, in the external ordering $(1234)$, is
\eqn\fullfour{\eqalign{\G(k_1, k_2, k_3, k_4)=&\d^{(d)}(\sum_ik_i)
\int_0^{\infty}{d\t\over \t^{{d\over 2}+1}}\t^4\int_0^{1}\prod_{i=1}^4 
d\a_i\d(\sum_i\a_i-1)\cr
\times &e^{-\t(\a_4\a_1k_1^2+\a_1\a_2k_2^2+ 
\a_2\a_3k_3^2+\a_3\a_4k_4^2+\a_2\a_4(k_1+k_2)^2+\a_1\a_3(k_1+k_4)^2)}.}}
We will expand the factor of $e^{-\t\a_1\a_3(k_1+k_4)^2}$ and take the fourier transform 
term by term. 

Let's look at the leading term in this expansion. We can do the Gaussian integral over the 
momenta after introducing a lagrange multiplier as in the case of the three point function. 
The presence of the term proportional to $(k_1+k_2)^2$ in the exponent of \fullfour\
suggests introducing  
another lagrange multiplier. That is we rewrite the momentum conserving delta function as 
\eqn\lagmult{\eqalign{\d^{(d)}(\sum_ik_i)=&\int d^dk_s\d^{(d)}(k_1+k_2-k_s)
\d^{(d)}(k_3+k_4+k_s)\cr
=&\int d^dk_sd^dzd^dwe^{i(k_1+k_2-k_s)\cdot z}e^{i(k_3+k_4+k_s)\cdot w}.}}
We now carry out the fourier transform with respect to the momenta $k_i$ and also perform the 
integral over $k_s$. Following steps similar to that in Sec. 3
we readily get the position space expression
\eqn\posfour{\eqalign{\G(x_1, x_2, x_3, x_4)=&\int_0^{\infty}{d\t\over \t^{{d\over 2}+1}}\t^4
\int_0^{1}\prod_{i=1}^4 d\a_i\d(\sum_i\a_i-1)\cr
\times &\int d^dzd^dw\la x_1|e^{\t\a_4\a_1\lform}|z\ra
\la x_2|e^{\t\a_1\a_2\lform}|z\ra \cr
\times & \la z|e^{\t\a_2\a_4\lform}|w\ra
\la w|e^{\t\a_2\a_3\lform}|x_3\ra\la w|e^{\t\a_3\a_4\lform}|x_4\ra .}}
Note that there are two intermediate positions $z, w$ 
that have to be integrated over. In this form we have clearly exhibited
the $s$-channel tree structure of fig. 4, in position space. 
Higher powers in the expansion of 
$e^{-\t\a_1\a_3(k_1+k_4)^2}$ can also be Fourier transformed in a similar way. Since the 
integrals are still Gaussian, the basic structure of \posfour\ persists. There are now 
multiplicative tensor structures in $x_i^{\mu}, z^{\mu}, w^{\mu}$ which are necessary for the 
description of the exchange of higher spin states, as well as conformal descendants. 

As we saw in the case of the three point function,
the above heat kernel structure of the tree was important for transforming 
the integrand into an $AdS$ amplitude. This was essentially because
the $AdS$ bulk to boundary propagators bore a close relation to
the $d$-dimensional heat kernel. 
To match terms as in \posfour\ to a 
contribution to the $s$-channel four point amplitude in $AdS$, the bulk-to-bulk propagator
in $AdS$ will also have to make an appearance. Since it too is a solution
to the wave equation in $AdS$ (albeit with a delta function source), it is not surprising that 
it can also be naturally written in terms of the heat kernel. 
For instance, the scalar bulk-to-bulk propagator in position space can be written in the 
Schwinger representation (see Appendix B)
\eqn\finbtob{G(z,w;t_1,t_2)=\sum_{n=0}^{\infty}{1\over n!\G({d-2\over 2}+n)}
{(4t_1t_2)^{{d-2\over 2}+n}\over (t_1+t_2)^{{d\over 2}-2+2n}} 
\int_0^{\infty}d\r\r^{{d\over 2}-3+2n}e^{-\r}\la z|e^{{t_1+t_2\over \r}\lform}|w\ra .}
where 
as before we have redefined $t_1=z_0^2, t_2=w_0^2$. 

This representation is a generalisation of \htk\  and is already in a suggestive form 
in relation to the worldline expressions.
Note that the $n=0$ term in \finbtob\ dominates as one of the 
bulk points approaches the boundary, and is proportional to the bulk to boundary propagator
in \htk . This is identified with the contribution of the conformal primary $\Tr\Phi^2$.  
The higher powers of $n$ can be identified with the 
contribution of spin zero conformal descendants $\lform^n\Tr\Phi^2$
of this operator \liu \dmmr\foot{There are however 
subtleties here involving logarithms \liu\ \dmmr .}. Similar representations 
exist for higher spin particles. So all the right ingredients are present for a match with the 
field theory.

What remains to be seen is that all these ingredients can be put 
together and an appropriate change of variables be made on the Schwinger parameter space 
so that the worldline expression 
goes over into an $AdS$ amplitude. Moreover, the intermediate states exchanged in any channel
have to be in
the leading Regge trajectory. We hope to verify this conjecture in detail in future
work. Our intention here has merely been to make it plausible 
to the reader that our considerations for the three point function generalise nontrivially 
to the fourpoint function.

\newsec{Final Remarks}   

We have taken some small steps here in trying to understand how free field theory 
could reorganise itself into a theory of closed string modes on a higher dimensional $AdS$ space.
Essentially, we have sought to carry through the logic of open-closed string duality.
We hope to have made the case that the worldline representation of the field theory 
is a natural framework within which this can be done. This was, perhaps, 
to be expected because it is the
appropriate limit of the open string. But, in addition, as we saw in Sec.5, 
there seems to be a systematic way in which 
the gluing of loops into tree structures takes place in this Schwinger parametrised 
representation. The analogy to electrical networks gives us the intuition as to how this happens. 

Of course, associating tree structures
to loops is just the geometrical aspect of the open-closed string duality. The dynamical aspect 
consists of understanding how, in this process, the background also changes from flat space to
$AdS$. Here, we do not yet have any systematic understanding. Nevertheless, the worldline
formalism has given some important clues in this direction. The close relation between 
propagators in $AdS$ and proper time propagators in the boundary theory is crucial for
the transmutation of the field theory amplitude into one on $AdS$. As we saw in the case of the two and
three point functions a fairly simple change of variables on the Schwinger parameters
takes us from one to the other. Though 
we have not yet worked out such a change of variables for the four point function, 
we believe it exists. Relatedly, the close identification
of the overall proper time with the radial coordinate appears to be some kind 
of realisation of ideas on the Liouville mode and the extra dimension. 
Therefore the worldline formalism seems to also have 
the power to manifest the change of 
background in the process of gluing loops into trees. 
We, however, need to go beyond a case by case change of variables and 
find a way to understand this in more generality. This will require some more insight into
the relation between the open and closed string parameters. Rather than working with some 
particular coordinatisation, as we have been doing, we perhaps need a more invariant 
characterisation of the respective moduli spaces. 

Actually, the entire discussion of the last paragraph is unavoidably tied up with the
issue of the closed string description of $AdS$\mettse\berk .\foot{It is understood 
that whenever we talk of a closed string theory on $AdS$, we mainly have in mind the maximally 
supersymmetric theory on $AdS_5\times S^5$.} We have been trying all along in this paper, to bypass 
this issue by restricting ourselves to the twist two operators. The idea, as mentioned in the introduction,
is that the dual description of this sector might conceivably only involve a point particle 
like limit of the string on $AdS$. Apart from the existence of a consistent classical 
theory of the massless higher spin fields, there is another source of intuition for this guess. 
All attempts that have been made to study closed strings on $AdS$ in the zero radius limit have 
found evidence for some kind of bit picture emerging from the closed string worldsheet \dmw\karch\polch .
These bits or partons are to be identified with the Yang-Mills fields. The bilinear operators  
are then those with the smallest allowed number of bits, namely two. We therefore
expect this case to be one 
where the worldsheet is slimmest and thus closest to that of a point particle. 
To fatten the worldsheet, we would need a large number of bits, 
as is familiar from usual lightcone considerations or, more pertinently, in the BMN picture \bmn . 
In any case, our computations, to their limited extent, seem to bear out this 
working hypothesis for the leading Regge trajectory.  

But even here it is clear that, even if the amplitudes 
can be viewed as those of point particles in $AdS$, it is a very cumbersome way of doing things.  
For instance, in the four point function, all the infinite number of particles in the leading Regge 
trajectory should appear as intermediate states. A sum over individual bulk-to-bulk $AdS$ propagators 
for all these states is not only technically demanding but also ugly. A look at Eq.\fullfour\ shows 
that expanding in the $s$-channel (i.e. in powers of $t$) mutilates a nice $s-t$ symmetric expression. 
It is the analogue of expanding the Veneziano amplitude in the $s$-channel which leads to messy
individual terms. Since the worldline expressions are in duality symmetric form, it should be possible 
to recast them directly into a duality symmetric closed string description. Perhaps the unbroken infinite 
dimensional higher spin symmetry on $AdS$ \vasrev\vas\sezsun\pol\witt\mikh\
should give us a hint on how to formulate such a 
description. After all, the free field Laplacian entering in the worldline formalism 
also has such a symmetry \east (see also \anselmi ).

Another clue should come from the generalisation to operators with more bits. As mentioned at the end of 
Sec. 5, correlation functions of operators like $\Tr\Phi^J$ for large $J$ will have many worldlines 
(and loops). It should be possible to examine the Schwinger parametrisation of these correlators 
and see an effective thickening of the worldsheet. It has recently been proposed \dnw\ that there is a
huge Yangian symmetry that acts on the set of all free partons which is related to the non-local
symmetries of the sigma model on $AdS$ \msw\bnp\vall . This would be a generalisation of the 
higher spin symmetries of the bilinears.

The idea of seeing an infinite number of unbroken symmetries in string theory in the limit 
of $\apm\R\infty$\foot{In the zero coupling limit, keeping 
the radius of $AdS$ fixed, like we have, is equivalent to taking $\apm\R\infty$.} 
goes back to Gross \gross\gm . Some of the 
features found in \gm , such as the contribution only of special kinds of worldsheets 
in high energy amplitudes, seem to have an echo in our considerations.\foot{We thank David Gross for 
discussions on these matters.} 

At nonzero coupling (or finite $\apm$) we expect these symmetries to be higgsed 
\witt\por\bianchi . The open-closed string duality would, nevertheless, 
continue to hold. We note, in this context,
that the electrical analogy holds for arbitrary Feynman diagrams, including that of an interacting
gauge theory. Hence we expect the gluing of loops into trees to be implemented in the worldline formalism
even at non-zero coupling. 
The generality of the worldline formalism might also be useful in trying to extend the 
open-closed string duality to nonsupersymmetric gauge theories. Perhaps this will also
enable us to tie these gauge string dualities with the ``other'' kind of open-closed duality
that takes place in the process of tachyon condensation (see \ashoke\ \gir\ for recent discussions).

\bigskip

\centerline{\bf Acknowledgements}
It is a pleasure to acknowledge the various helpful conversations I have had over the months 
with A. Adams, O. Aharony, S. Cherkis,
A. Dhar, S. Das, J. R. David, M. Douglas, 
E. Gimon, D. Ghoshal, D. J. Gross, S. Govindarajan, F. Hassan, D. Jatkar, 
S. D. Joglekar, S. Kachru, H. Liu, J. Maldacena,
G. Mandal, E. Martinec, 
S. Minwalla, L. Motl, H. Neuberger, A. Petkou, 
M. Rangamani, S-J. Rey, K. Schalm, A. Sen, E. Silverstein, K. Skenderis,
A. Strominger, S. P. Trivedi, C. Vafa, P. Windey, K. P. Yogendran and M. Zamaklar.  
I must specially extend my thanks to 
I. Klebanov, L. Rastelli, S. Wadia and E. Witten for several
discussions, remarks and general encouragement 
which helped shape the course of this investigation. I am also grateful to A. Sen and S. Wadia
for comments on the manuscript. 
A large part of this work was carried out 
while being a visiting member at the Institute for Advanced Study
where the author's work was supported by 
DOE grant DE-FG02-90ER40542.  
The hospitality of the physics departments 
at Harvard University, Rutgers University and I.I.T. Kanpur is also gratefully acknowledged. 
As also that of the organisers of the
Amsterdam Workshop on String Theory and the Crete Regional Meeting 
in String Theory, where preliminary versions of this work were presented.
Most of all, thanks are due to the people of India for their unstinting support. 

\appendix{A}{The Two Point Function} 

The two point function is a simple illustration of the ideas in the main body of the paper. 
The worldline expression for the two point correlator is given from Eq.\wlint\ to be, after 
some obvious change of variables 
\eqn\twopt{\G(k_1, k_2)=\d^{(d)}(k_1+k_2)
\int_0^{\infty}{d\t\over \t^{{d\over 2}+1}}\t^2\int_0^1 d\a 
e^{-\t\a(1-\a)(\b k_1^2+(1-\b)k_2^2)},}
where $\b$ is arbitrary since the integrand is actually independent of it. 
This is more conventionally written in terms of the reduced form 
\eqn\twosim{\tilde{\G}(k)=
\int_0^{\infty}{d\t\over \t^{{d\over 2}+1}}\t^2\int_0^1 d\a 
e^{-\t\a(1-\a)k^2}.}
Eq. \twosim\ is a straightforward example of the gluing up process and its interpretation
in terms of the electrical analogy. The loop with two insertions is glued up into a tree which is
just a line segment in this case. In electrical terms, this is just the elementary fact that
two parallel resistors (in this case proportional to the parameters $\a$ and $(1-\a)$) can be
replaced by a single equivalent resistor (proportional to $\a(1-\a)$ for us). This is evident 
from the exponent in \twosim .

As before, the position space expression corresponding to \twopt ,\twosim\ is 
\eqn\postwo{\eqalign{\G(x_1, x_2)
=&\int_0^{\infty}{d\t\over \t^{{d\over 2}+1}}\t^2\int d^dz\int_0^1 d\a
\la x_1|e^{\t\a(1-\a)\b\lform}|z\ra\la z|e^{\t\a(1-\a)(1-\b)\lform}|x_2\ra \cr
=&\int_0^{\infty}{d\t\over \t^{{d\over 2}+1}}\t^2\int_0^1 d\a
\la x_1|e^{\t\a(1-\a)\lform}|x_2\ra .}}
As in the case of the three point function, the position space expression clearly exhibits the 
glued up form of the loop.
We will relate \postwo\ to the two point amplitude in $AdS$. 

The latter is essentially proportional to the convolution of two bulk-to-boundary propagators
\foot{We are being a little cavalier here.
Actually, there are contributions from gradient terms as well. 
However, because of the equation of motion, these are related to each other and 
one is left with a boundary term which needs to be treated carefully \fmmr . 
Since our interest is not in reproducing the right normalisation factors, but 
rather in seeing how loops glue into $AdS$ trees, it suffices to consider 
the product of two bulk-to-boundary propagators. The price 
we will pay is that some expressions will be formally divergent.}
so that just as in \threetree\ we have 
\eqn\twoads{\G(x_1, x_2)=\int d^dz\int_0^{\infty}{dt\over t^{{d\over 2}+1}}t^2
\int_0^{\infty}d\r_1d\r_2(\r_1\r_2)^{{d\over 2}-3}e^{-\r_1-\r_2} 
\la x_1|e^{{t\over 4\r_1}\lform}|z\ra\la z|e^{{t\over 4\r_2}\lform}|x_2\ra.}
A change of variables to $\r_1=\r(1-\b), \r_2=\r\b$ and introducing a trivial integral over $\a$
makes this take the form
\eqn\twob{\eqalign{\G(x_1, x_2)=&\int d^dz\int_0^{\infty}{dt\over t^{{d\over 2}+1}}t^2
\int_0^{\infty}d\r\r^{d-5}e^{-\r}\int_0^1d\b\int_0^1d\a [\b(1-\b)]^{{d\over 2}-3}\cr
\times &\la x_1|e^{{t\over 4\r(1-\b)}\lform}|z\ra
\la z|e^{{t\over 4\r\b}\lform}|x_2\ra.}}
We can now relate $t$ to the proper time $\t$ through
\eqn\ttau{t=4\t\r\b(1-\b)\a(1-\a).} 
Note the similarity to Eq.\proptime .
The integral over $\r$ decouples and we are left with
\eqn\twofin{\eqalign{\G(x_1, x_2)=&
\int_0^1 {d\b\over \b(1-\b)}\int_0^{\infty}{d\t\over \t^{{d\over 2}+1}}\t^2 \int d^dz
\int_0^1{d\a\over [\a(1-\a)]^{{d\over 2}-2}}\cr
\times &\la x_1|e^{\t\a(1-\a)\b\lform}|z\ra
\la z|e^{\t\a(1-\a)(1-\b)\lform}|x_2\ra \cr
=&\int_0^1 {d\b\over \b(1-\b)}\int_0^{\infty}{d\t\over \t^{{d\over 2}+1}}\t^2
\int_0^1{d\a\over [\a(1-\a)]^{{d\over 2}-2}}\la x_1|e^{\t\a(1-\a)\lform}|x_2\ra .}}
Modulo the overall divergent factor from the decoupled 
$\b$ integral, \twofin\ coincides with Eq.\postwo\ for $d=4$, the case of interest.
The overall divergence is not unexpected given the comments in the previous footnote.

\appendix{B}{The Scalar Bulk to Bulk Propagator}

The position space bulk-to-bulk propagator in $AdS$ for a scalar field 
corresponding to an operator of dimension $\D$ is usually put in the form (see for e.g. \liu ) 
\eqn\btob{G(z,w;z_0,w_0)=\xi^{-\Delta}F({\D+1\over 2}, {\D\over 2}, \D-{d\over 2}+1;{1\over \xi^2}),}
where 
\eqn\defxi{\xi={z_0^2+w_0^2+|z-w|^2\over 2z_0w_0}.}  
In the case of the free field operator $\Tr\Phi^2$
with $\D=d-2$, the hypergeometric function in \btob\ 
simplifies and
\eqn\simpbtob{G(z,w;z_0,w_0)=\xi^{-\Delta}{1\over (1-{1\over \xi^2})^{d-1\over 2}}.}
Using the expansion 
\eqn\hypexp{{1\over (1-z)^a}=\sum_{n=0}^{\infty}{\G(n+a)\over \G(a)}{z^n\over n!}.}
and redefining $z_0^2=t_1, w_0^2=t_2$, we can write the bulk-to-bulk propagator in
a Schwinger parameter expansion very similar to Eq.\htk .
\eqn\schbtob{\eqalign{G(z,w;t_1,t_2)=&\sum_{n=0}^{\infty}{\G(n+{d-1\over 2})\over n!\G(d+2n-2)}
{(4t_1t_2)^{{d-2\over 2}+n}\over (t_1+t_2)^{d-2+2n}}\int_0^{\infty}d\r\r^{d-3+2n}e^{-\r}
e^{-\r{|z-w|^2\over t_1+t_2}}\cr
=&\sum_{n=0}^{\infty}{1\over n!\G({d-2\over 2}+n)}
{(t_1t_2)^{{d-2\over 2}+n}\over (t_1+t_2)^{{d\over 2}-2+2n}} 
\int_0^{\infty}d\r\r^{{d\over 2}-3+2n}e^{-\r}\la z|e^{{t_1+t_2\over 4\r}\lform}|w\ra .}}
Here we have used the Schwinger representation
\eqn\gam{{1\over \xi^{\l}}={1\over \G(\l)}\int_0^{\infty} d\r\r^{\l-1}e^{-\r\xi}.}
as well as the identity
\eqn\gamid{\G(\D+2n)={1\over 2\pi^{\half}}2^{\D+2n}\G({\D+1\over 2}+n)\G({\D\over 2}+n).}

\appendix{C}{Divergences and the UV-IR Relation}

Let us consider the free field theory in an arbitrary curved background (or more generally one in 
which some of the higher spin fields also have an expectation value). The action is still quadratic
but the effective action is now a complicated nonlocal functional of the metric (and other fields). 
There are some UV divergences in this effective action, but they are local in the 
background fields. This is familiar from the study of quantum fields in curved space.
The conventional way to isolate these divergences is, in fact, the heat kernel or proper time expansion. 
This involves studying the proper time representation of the effective action in the background and 
putting a UV cutoff $\ep$ at small proper times to regularise the expression. We then make a 
small time expansion and isolate the leading divergent pieces.  

What we would like to remark here is that the same structure of divergences is present in the 
effective action on $AdS$ as a functional of the boundary values of the metric (and other fields). 
The difference is that these divergences are now in the IR and can be regularised by an IR cutoff 
$\epp$ in the radial coordinate of $AdS$. This fits in well with our picture where the proper time 
essentially transmutes itself into the radial coordinate of $AdS$. 

\subsec{The Heat Kernel Expansion}

The effective action for, say, the free adjoint scalar field in a curved background $h_{\mu\nu}$
is given in a heat kernel representation. 
\eqn\effac{{1\over N^2}\G(h_{\m\n})=\half\ln \rm{det}(-\lform_h)
=\half\int_{\ep}^{\infty}{d\t\over \t}\Tr[e^{\t\lform_h}].}
where we have put in the UV cutoff $\ep$ and indicated the curved background in the 
subscript for the Laplacian. For the fermions and gauge fields there are analogous representations of the
corresponding kinetic operators. 
For small proper times the trace of the heat kernel has the well known Schwinger-Dewitt expansion in 
terms of local functionals (see \birdav\ for instance)
\eqn\dws{\Tr[e^{\t\lform_h}]\sim\int {d^dz\sqrt{h}\over \t^{d\over 2}}[\sum_{j=0}^{[{d\over 2}]}
a_j(z)\t^j+ \ldots],}
where the $\ldots$ indicate terms which give UV finite contributions. The sum over $j$ is a derivative 
expansion. The $a_j(z)$ are the familiar Schwinger-DeWitt coefficents which are curvature
invariants built from $h_{\m\n}$ and having a total of $2j$ derivatives of the metric. 
Thus (with the appropriate normalisation which we have omitted) $a_0=1$, $a_1(x)={1\over 6}R$ etc.

Therefore, the effective action has the expansion
\eqn\effexp{{1\over N^2}\G(h_{\m\n})=\int_{\ep}^{\infty}{d\t\over \t^{{d\over 2}+1}}\int d^dz\sqrt{h}
[1+{\t \over 6}R+\t^2 O(R^2)+\ldots].}
Not accidentally, this is like the worldline expressions we had for correlation functions. 
In even dimensions the term with $j={d \over 2}$ in \dws\ has a logarithmic dependence on $\t$ and 
gives rise to a logarithmically divergent term which is 
responsible for the conformal anomaly.

\subsec{Comparison to $AdS$}

The measure in \effexp\ has the right structure to be that of $AdS_{d+1}$. In fact, the form of the 
integrand is also what one would have for a classical action on $AdS$ evaluated onshell. 
When we speak here of a classical action on $AdS$ we do not have in mind some kind of Einstein-Hilbert
action or supergravity variant. It could be more like a string field 
action involving an infinite number of derivatives as in
the Vasiliev theories. 

But already at the level of the Einstein-Hilbert action, one sees a very similar structure, on-shell, 
to \effexp .
One solves 
Einstein's equations on $AdS_{d+1}$ with an asymptotic boundary metric $h_{\m\n}(z)$ by parametrising 
the bulk metric to be 
\eqn\adsmet{ds^2={dt^2\over t^2}+{h_{\m\n}(z,t)dz^{\m}dz^{\n}\over t} ,}
where as $t\R 0$
\eqn\metasym{h_{\m\n}(z,t)\R h^{(0)}_{\m\n}(z)+t h^{(2)}_{\m\n}(z)+t^2h^{(4)}_{\m\n}(z) +\cdots +
t^{[{d\over 2}]}h^{(d)}_{\m\n}(z)+t^{[{d\over 2}]}\ln{t}\tilde{h}^{(d)}_{\m\n}(z)+\cdots .}
Here $h^{(0)}_{\m\n}(z)=h_{\m\n}(z)$.
The observation of Fefferman and Graham \feffgr\ was that one can solve for $h^{(2)}_{\m\n}(z), 
h^{(4)}_{\m\n}(z), \ldots \tilde{h}^{(d)}_{\m\n}(z)$ (but not $h^{(d)}_{\m\n}(z)$) algebraically
in terms of $h^{(0)}_{\m\n}(z)$. Putting this back into Einstein's equations, gives \hsken\  
\eqn\adsexp{{1\over N^2}\G(h_{\m\n})=\int_{\epp}^{\infty}{dt\over t^{{d\over 2}+1}}\int d^dz\sqrt{h}
[\sum_{j=0}^{[{d\over 2}]} \tilde{a}_j(z)t^j+ \ldots].}
Here $\tilde{a}_j(z)$ are local curvature invariants of dimension $2j$ built from $h_{\m\n}(x)$ 
just as $a_j(z)$ in \dws . In general, $\tilde{a}_j(z)$ and $a_j(z)$ are distinct linear combinations 
of the finite number of curvature invariants of dimension $2j$. 
But Henningson and Skenderis \hsken\ could compare
the conformal anomaly of the field theory ($j={d\over 2}$ piece of \dws ) 
with the similarly logarithmically divergent ($j={d\over 2}$) piece of \adsexp .
For the full $\CN=4$ Yang-Mills multiplet, they found precise agreement. See also \nojod . 

Our purpose here is just to point out that
the similarity between \dws\ and \adsexp\ is another 
signature of the role of 
the proper time representation of the field theory in reconstructing the bulk  
description. The connection between the 
radial coordinate and the the propertime  shows up clearly over here\foot{It is interesting that
the natural parametrisation for the $AdS$ metric \adsmet\ in \feffgr\ \hsken\ 
employs $t$ rather than the more common
$z_0^2=t$.}. In the examples that we have seen, such as \proptime\ and \ttau , the two have
been multiplicatively related to each other. Thus, though the cutoffs $\ep$ and $\epp$ cannot be 
directly identified with each other, because of their multiplicative relation
the logarithmically divergent pieces can be compared. This is also related to 
the fact that the power law divergent terms
for $j<{d\over 2}$ are prescription dependent in the field theory whereas the logarithmically 
divergent one isn't.    

Similar calculations including backgrounds for scalar fields in $AdS$ have been carried out in 
\odnoj\dss\ 
with very similar results to the corresponding heat kernel expansion in field theory. 
Perhaps utilising the higher spin symmetries of the free laplacian \east\ one might be able
to relate the heat kernel in an arbitrary quadratic background 
to on-shell actions of the Vasiliev type, generalising the remarks in this section.

\listrefs

\end